\documentclass[a4paper,12pt]{article}

\usepackage[hmargin = 2cm, vmargin = 2cm]{geometry} 

\usepackage{authblk} 
\usepackage{titling}
\usepackage{amsthm}
\usepackage{thmtools}
\newtheorem{theorem}{Theorem}
\declaretheorem[name=Theorem]{reptheorem}
\usepackage{url}
\usepackage{amsmath, amssymb}
\usepackage{graphicx}
\usepackage{bm}  
\usepackage{tikz, pgfplots}
\pgfplotsset{compat=1.18}
\usepackage{color, xcolor}
\usepackage[table]{xcolor}
\newcommand{\cg}{\cellcolor[HTML]{C0C0C0}}
\newcommand{\cb}{\color[HTML]{3166FF}}
\usepackage{float}
\usepackage{natbib}
\usepackage{rotating}
\usepackage{pdflscape}
\usepackage[normalem]{ulem}
\usepackage{listings}
\usepackage[doublespacing]{setspace}
\usepackage[style=british]{csquotes}  
\usepackage{multirow, multicol, adjustbox}
\usepackage{booktabs}  
\usepackage{tabularx}  
\usepackage{siunitx}  
\usepackage{subcaption} 
\usepackage{algorithm}
\usepackage{algpseudocode}
\usepackage{pgfgantt}
\usepackage{array}
\usepackage{pdfpages}
\usepackage[font=scriptsize]{caption}
\usepackage{titlesec}
\usepackage{threeparttable}
\usepackage{appendix}
\usepackage{hyperref}
\hypersetup{
    colorlinks=true,
    linkcolor=blue, 
    citecolor=blue, 
    filecolor=magenta, 
    urlcolor=blue 
}

\newcounter{gaocomm}
\newcounter{Note}
\definecolor{blue-violet}{rgb}{0.00,0.75,0.90}
\definecolor{mygreen}{rgb}{0.0, 0.5, 0.0}
\definecolor{awesome}{rgb}{1.0, 0.13, 0.32}
\definecolor{bostonuniversityred}{rgb}{0.8, 0.0, 0.0}

\renewcommand\arraystretch{1.4}

\title{\Large Graph Signal Processing for Global Stock Market Realized Volatility Forecasting}
\author{Zhengyang CHI\thanks{Corresponding author. Email: zhengyang.chi@sydney.edu.au}}
\author{Junbin GAO}
\author{Chao WANG\thanks{Email for all authors: \{zhengyang.chi, junbin.gao, chao.wang\}@sydney.edu.au}}
\affil{\small Discipline of Business Analytics, The University of Sydney Business School}
\begin{document}
\date{}
\maketitle
\begin{abstract}
This paper introduces an innovative realized volatility (RV) forecasting framework that extends the conventional Heterogeneous autoregressive (HAR) model via integrating Graph Signal Processing (GSP). The study first evaluates various constructions of volatility-interrelationship networks by analyzing how the associated graph signal energy tracks global financial market volatility. Volatility spillovers are subsequently embedded into the proposed framework, which employs the graph Fourier transform (GFT) and its inverse to effectively capture global stock market dynamics in both the spectral and spatial domains. The framework not only provides a global context for modeling the volatility interrelationships, but also captures the nonlinearity and directionality of the volatility spillover effect. The empirical study using RV data of $24$ global stock market indices compares short-, mid- and long-term RV forecasts with various HAR-type benchmarks and a graph neural network-based HAR model. The proposed model consistently outperforms all comparators, demonstrating the effectiveness of integrating GSP into the HAR model for RV forecasting.
\end{abstract}


\newpage
\pagenumbering{arabic}

\section{Introduction}
In the complex ecosystem of global financial markets, understanding and forecasting volatility is critical for investors, risk managers and policymakers. To capture the dynamics of financial volatility, numerous measures have been proposed \citep{Poon2003}. Among these, the high-frequency-data-based Realized Volatility (RV) \citep{Andersen1998} has become a standard tool in both research and practice. It is widely used for forecasting, risk management and asset pricing.

Volatility is rarely confined to individual assets or markets-it propagates across financial systems via spillovers, contagion and co-movements, especially during crises. Recent studies on financial volatility analysis have highlighted volatility spillover phenomena, with interest intensifying in the wake of the 2008 Global Financial Crisis (GFC) and the COVID-19 pandemic. This phenomenon refers to the transmission of volatility shocks from one market (or asset) to others \citep{Kanas2000, Forbes2002, Poon2003, Diebold2009, Yang2017, Bollerslev2018}. The significance of the volatility spillover effect necessitates a comprehensive volatility forecasting framework that jointly captures the cross-sectional volatility interconnections across different financial markets and the temporal autocorrelation of each market's volatility. Recent work has applied graph-based models to RV forecasting by exploiting spillover networks \citep{Zhang2025}, demonstrating that explicitly specifying the structure of inter-market linkages can enhance the RV forecast accuracy.

Yet, an important aspect remains largely unexplored in the graph-based RV forecasting studies: the lack of a systematic way to evaluate how well a given network construction represents the true interdependence structure. Since different graph-building methods (e.g., correlation-based, variance decomposition–based, or directional measures) can yield vastly different network topologies, the absence of an evaluation framework, which is agnostic to volatility forecasting models, makes it difficult to judge which constructions meaningfully represent volatility dynamics. Without such diagnostics, forecasts may rely on networks whose structure is potentially misaligned with the underlying economic reality, reducing interpretability and performance. To bridge this gap, we introduce graph signal energy (GSE) as a novel diagnostic tool to assess the informativeness or effectiveness of different network constructions before they are embedded into forecasting models. The concept builds on graph signal processing (GSP), a rapidly growing field that generalizes classical signal processing to graph-structured data \citep{ortega2018graph}. In GSP, the value of a variable observed on each node is treated as a graph signal, while graph edges encode relationships or dependencies among nodes. The GSE then quantifies how ``smooth'' or ``irregular'' this signal is with respect to the graph topology, based on the graph Laplacian quadratic form. 

\begin{figure}[H]
    \centering
    \includegraphics[width=15cm]{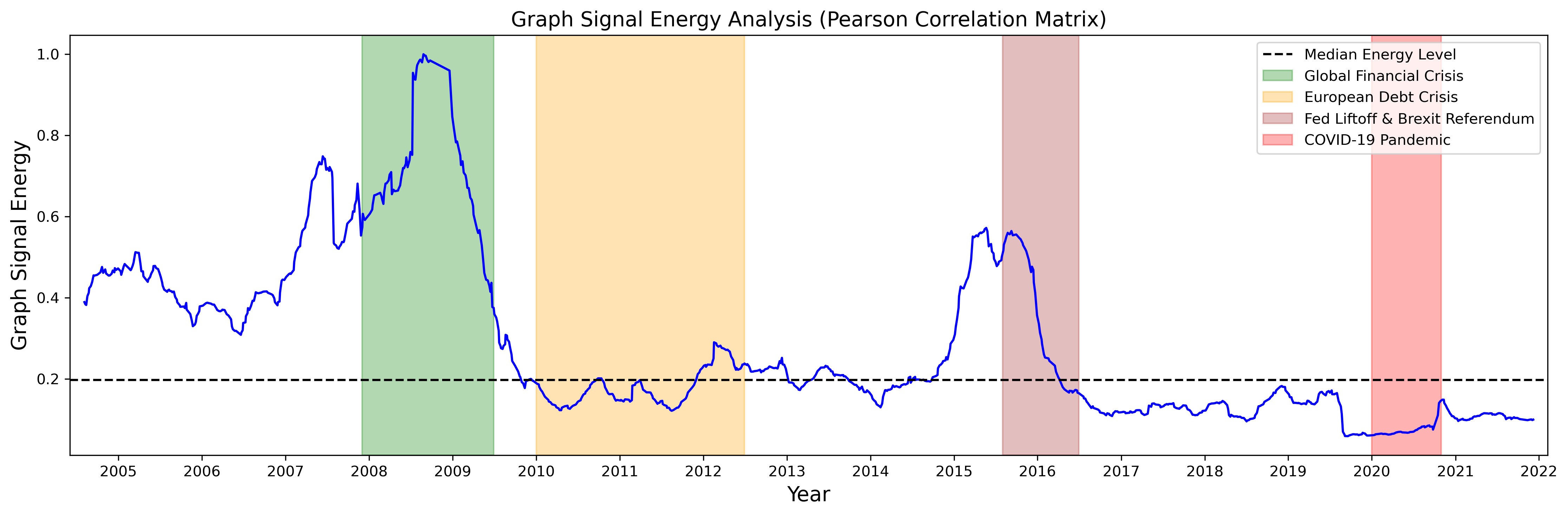}
    \caption{The RV GSE of the volatility spillover network based on the Pearson correlation matrix}
    \label{graph_energy_Pearson}
\end{figure}

\begin{figure}[H]
    \centering
    \includegraphics[width=15cm]{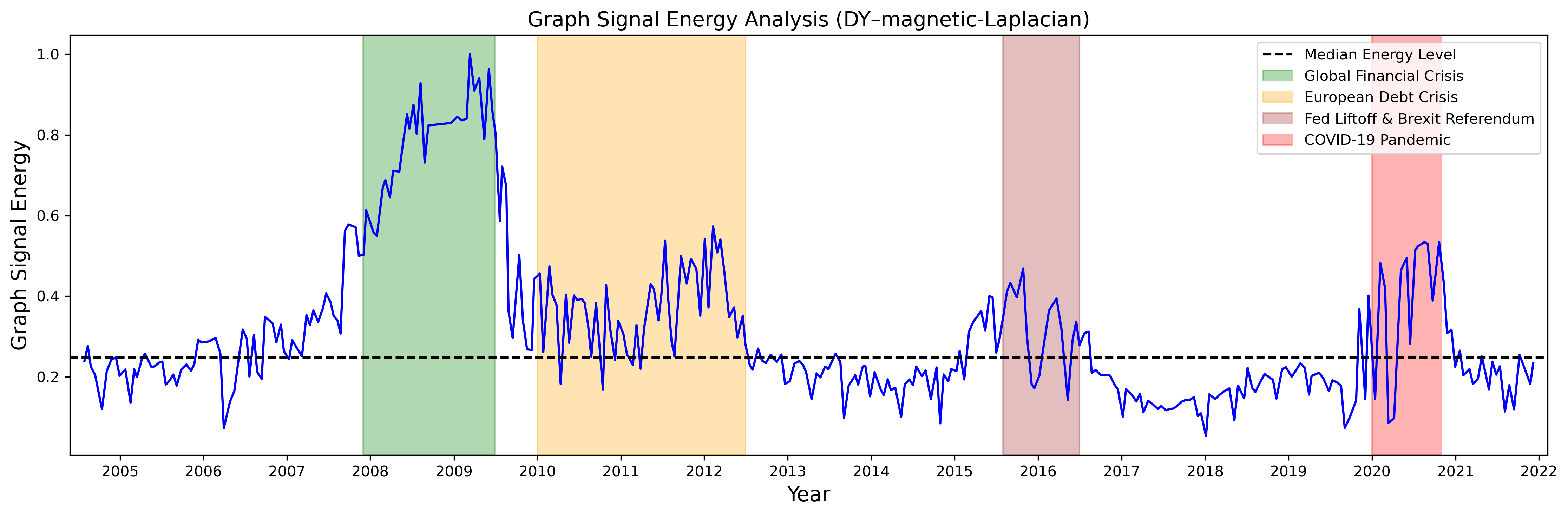}
    \caption{The RV GSE of the volatility spillover network based on the DY-magnetic-Laplacian construction method}
    \label{graph_energy_DY_magnet}
\end{figure}

Applied to RV spillover networks in a rolling-window setting, GSE provides an intuitive and theoretically grounded way to assess whether a network topology is informative for capturing volatility interdependencies. Especially, we expect RV GSE to behave differently in turbulent versus stable periods because the strength of cross-market volatility linkage varies across different market conditions \citep{Diebold2012, Bensaida2018}. For illustration, we compare RV GSE patterns for networks constructed from the Pearson correlation matrix and the Diebold-Yilmaz (DY) framework \citep{Diebold2012} combined with the magnetic Laplacian matrix \citep{Zhang2021}. In Figure \ref{graph_energy_Pearson}, the RV GSE of the Pearson correlation–based network remains unexpectedly low during major volatility spikes such as the European Debt Crisis and the COVID-19 pandemic, but reaches a relatively high level in the comparatively stable year of 2005. In contrast, Figure \ref{graph_energy_DY_magnet} shows that the RV GSE of the DY-based network with the magnetic Laplacian exhibits the expected dynamics. It rises during turbulent periods, including the GFC, European Debt Crisis, and COVID-19, and remains subdued in tranquil periods. These findings demonstrate that GSE provides a powerful, model-agnostic lens for evaluating graph construction methods, and suggest that the DY-magnetic-Laplacian construction method offers a more informative and theoretically consistent representation of volatility spillover networks than the Pearson correlation approach. Further details of the GSE computation and its integration into volatility spillover analysis with theoretical justifications are provided in Sections \ref{Graph Theory Fundamentals} and \ref{Volatility Spillover Network and Its GSE}.

Having identified the DY-magnetic-Laplacian method yields a more informative representation of the RV spillover network, this study next employs GSP techniques to fully exploit the topological advantages of the chosen graph. GSP provides a principled framework for analyzing and forecasting signals defined over networks, making it particularly well-suited to capture the cross-market volatility dynamics embedded in the spillover structure. The integration of the magnetic Laplacian with GSP methods is not merely a technical choice, but a necessity: it allows directional spillovers to be incorporated while ensuring that the spectral properties of the network are leveraged in the modeling process, which is ideal for the chosen DY-magnetic-Laplacian construction method. The methodological rationale is discussed in detail in Section \ref{Methodology and Implementation}.

Building on this foundation, we propose a new GSP-based forecasting method that extends the well-established Heterogeneous Autoregressive (HAR) framework for RV forecasting. The HAR model, introduced by \citet{Corsi2009} and subsequently extended by \citet{patton2015good}, \citet{Bollerslev2018} and others, has remained popular in financial econometrics due to its balance of simplicity and empirical effectiveness. By embedding HAR dynamics into a GSP framework (GSP-HAR), this study aims to enhance the multi-asset RV forecasting accuracy by explicitly accounting for volatility interdependencies across markets, which is not captured by the traditional HAR models.

Therefore, this paper makes two main contributions to the literature on RV forecasting. Firstly, this study employs the GSE as a diagnostic tool for evaluating the effectiveness of different volatility spillover network constructions. By examining how well the time-varying behavior of RV GSE aligns with financial market conditions, the analysis provides an intuitive and model-agnostic way to verify whether a network topology meaningfully reflects volatility interdependencies. This approach highlights a straightforward yet often overlooked principle: more accurate volatility forecasts can be achieved by first identifying the most informative network construction and then tailoring graph-based models to that structure.

Secondly, the research introduces the GSP-HAR framework, integrating GSP into the well-established HAR model. While existing HAR extensions incorporate volatility spillover effects by adding intermarket variables and thereby treating volatility linkages as exogenous covariates \citep{Bubak2011, Liang2020, Zhang2025}, the GSP-HAR framework instead embeds the volatility spillover effects within the graph spectral domain. By combining the volatility interrelationships and historical RV patterns, the proposed approach explicitly models both temporal and cross-market volatility dynamics. Empirical results demonstrate that this integration not only captures volatility spillovers more effectively, but also delivers substantial improvements in RV forecasting accuracy.                                                                

The paper is organized as follows. Section \ref{Background} provides preliminaries of pertinent concepts and models. The relevant volatility GSE calculation, the application of the magnetic Laplacian and GSP techniques and the proposed framework are developed in Section \ref{Methodology and Implementation}. Section \ref{sec: Experiments} presents the conducted experiments and their results. The implementation code is available at {\small\url{https://github.com/MikeZChi/GSPHAR.git}}. Section \ref{sec: Conclusion} concludes the paper.

\section{Background}
\label{Background}
Before presenting the construction of the volatility spillover network and the proposed GSP-HAR model, we first review the fundamental components of graph theory along with the commonly used measures and representations of graph topological structure. These include the weight matrix, the degree matrix and the Laplacian matrix. Particularly, the Laplacian matrix serves as a key tool for calculating the GSE. Additionally, we briefly review the HAR model and its volatility spillover extensions, which are commonly employed in RV forecasting.

\subsection{Graph Theory Fundamentals}
\label{Graph Theory Fundamentals}
A graph can be denoted as $\mathcal{G} = (\mathcal{V}, \mathcal{E})$, where $\mathcal{V}$ is the node set and $\mathcal{E}$ is the edge set of the graph. An edge $e_{ij} \in \mathcal{E}$ can be formulated as $e_{ij} = (v_i, v_j)$, which indicates that the edge $e_{ij}$ starts from node $v_i$, and ends at node $v_j$. In general, we consider the graph as a directed graph, thus $e_{ij} \neq e_{ji}$. However, if $e_{ij} = e_{ji}$ for all $i,j$, the graph is undirected. Examples of undirected and directed graphs can be found in Appendix \ref{Undirected vs Directed Graphs}. In a directed graph, two types of neighbors can be formulated for a node $v_i$. The outgoing neighbors are all nodes $v_j$ that $v_i$ points to: $N_i^{out} = \{v_j \ | \ e_{ij} \in \mathcal{E}\}$. Besides, the incoming neighbors are all nodes $v_j$ that point to $v_i$ points: $N_i^{in} = \{v_j \ | \ e_{ji} \in \mathcal{E}\}$.

Edges can have weights to represent the importance or strength of the connection. The weight matrix $\mathbf{W}$ is used to capture the structure of graphs. Suppose the weight attached to edge $e_{ij} \in \mathcal{E}$ is a positive real number $w_{ij}$. If the graph $\mathcal{G}$ has $N$ nodes, the formulation of the weight matrix $\mathbf{W} \in \mathbb{R}^{N \times N}$ is presented below:
\begin{equation} \label{directed weighted adj_mx}
    \mathbf{W}_{ij} =
    \begin{cases}
    w_{ij}, &\mbox{$ e_{ij} \in \mathcal{E}$};\\
    0, &\mbox{$e_{ij} \notin \mathcal{E}$}.
    \end{cases}
\end{equation}
For an undirected graph, its weight matrix is always symmetric because if $e_{ij} \in \mathcal{E}$, $e_{ji} = e_{ij}$ and, thus, $w_{ji} = w_{ij}$. On the other hand, the weight matrix of a directed graph is asymmetric. The degree matrix $\mathbf{D}$ is formulated based on the weight matrix as $\mathbf{D} = \text{diag}(d_1, \ldots, d_N)$, where $d_i = \sum_{j = 1}^{N} \mathbf{W}_{ij}$ measures the total connection strength of node $v_i$ to all others.

Besides, the Laplacian matrix of the graph is formulated as $\mathbf{L} = \mathbf{D} - \mathbf{W}$. Furthermore, the Laplacian matrix can be normalized as $\mathbf{L}_{norm} = \mathbf{I} - \mathbf{D}^{-\frac{1}{2}}\mathbf{W}\mathbf{D}^{-\frac{1}{2}}$ to reduce the influence of node degree variations, and become more numerically stable. To ensure that the degree matrix $\mathbf{D}$ is invertible, all its diagonal elements $d_i (i = 1, \ldots, N)$ must be positive. This can be easily achieved by assuming that the graph contains no isolated nodes. This assumption is reasonable in the real financial world because volatility is seldom limited to individual assets or markets. Since ensuring the invertibility of the degree matrix is straightforward, the following discussions adopt the normalized Laplacian matrix, and the subscript $_{norm}$ is omitted for notational simplicity.

First, we consider an undirected graph $\mathcal{G}$, consisting of $N$ nodes with no isolated nodes, a symmetric weight matrix $\mathbf{W}$ with non-negative entries, and an invertible degree matrix $\mathbf{D}$. A graph signal $x$ is defined as a mapping from $\mathcal{V}$ to $\mathbb{R}$ for each node in the graph. Given the finite cardinality $N$ of $\mathcal{V}$, the graph signal can be regarded as the collection of signal values observed at the nodes of the graph, and it is represented as a vector $\mathbf{x} \in \mathbb{R}^{N}$, where each component corresponds to the node value under specific fixed node indexing. The GSE $E(\mathbf{x}) \in \mathbb{R}$ quantifies how much the signal changes across connected nodes. It is calculated as:
\begin{equation} \label{graph_signal_energy}
    E(\mathbf{x}) = \mathbf{x}^\top \mathbf{L} \mathbf{x}.
\end{equation}

\begin{theorem}\label{Non-negative GSE theorem}
For any given undirected and non-negative weighted graph $\mathcal{G}$ with no isolated nodes, i.e. $\mathbf{W} = \mathbf{W}^\top$, $\mathbf{W}_{ij}\geq 0$ $(i,j=1, ..., N)$, and $d_i>0$ $(i=1, 2, ..., N)$, any graph signal $\mathbf{x} \in \mathbb{R}^N$ on $\mathcal{G}$ has non-negative energy: $E(\mathbf{x}) \geq 0$.
\end{theorem}
Thus, the GSE is capable of measuring the lack of smoothness of the graph signal $\mathbf{x}$ over the graph structure. The proof of Theorem \ref{Non-negative GSE theorem} is provided in Appendix \ref{Non-negative GSE}. 

\subsection{HAR Model and Its Extensions on Volatility Spillovers}
\label{HAR Model and Its Extensions on Volatility Spillovers}
The HAR framework is designed to account for the fact that different market participants, such as quantitative traders, institutional investors and long-term investors, operate on different time horizons. The HAR model incorporates pooled RV data from these various time scales, typically daily, weekly and monthly, to represent the volatility behaviors. This design helps to capture the heterogeneous nature of volatility patterns in financial markets.

Suppose $\mathbf{v}_t$ contains RV observations of $N$ stock market indices on common trading day $t$: $\mathbf{v}_t = (\text{RV}_{1,t}, \text{RV}_{2,t}, \ldots, \text{RV}_{N,t})^\top$. For integers $n > 1$, $\mathbf{v}_{t-n:t-1}$ is the average-pooled data: $\mathbf{v}_{t-n:t-1} = \frac{1}{n-1}\sum_{j = t-n} ^ {t-1} \mathbf{v}_j$. For notational simplicity, set the past daily, weekly and monthly RV pattern as $\mathbf{v}_d = \mathbf{v}_{t-1}$, $\mathbf{v}_w = \mathbf{v}_{t-5:t-1}$ and  $\mathbf{v}_m = \mathbf{v}_{t-22:t-1}$. $\widehat{\mathbf{v}}_t$ denotes the forecast RV value on day $t$. The conventional HAR model is formulated as:
\begin{equation} \label{HAR}
    \widehat{\mathbf{v}}_t = \boldsymbol{\alpha}  + \boldsymbol{\beta}_{d} \mathbf{v}_d + \boldsymbol{\beta}_{w} \mathbf{v}_w + \boldsymbol{\beta}_{m} \mathbf{v}_m,
\end{equation}
where coefficients $\boldsymbol{\beta}_{d}$, $\boldsymbol{\beta}_{w}$ and $\boldsymbol{\beta}_{m}$ are diagonal matrices of size $N$ and they measure how the past daily, weekly and monthly volatility patterns influence future RVs. The intercept term $\boldsymbol{\alpha}$ can be a vector. The HAR model cannot capture the comovement or correlations of the volatility in different stock markets, which limits its ability to produce more accurate forecasts.

The GNN-HAR model adds a $L$-layer spatial Graph Neural Network (GNN) component to the HAR model as an additional input variable to capture the volatility spillover effect in a nonlinear way. The GNN component relies on a precomputed volatility relational network $\mathcal{G}$ where the nodes are markets (or assets) and the edges are the volatility linkages. The weight matrix $\mathbf{W}$ is provided and remains fixed during the GNN-HAR training process. The GNN component aggregates past daily, weekly, and monthly volatility patterns by degree-weighted averaging over the neighborhood of each node (market), which is defined by the volatility spillover network. This aggregation is followed by learnable linear and non-parametric non-linear transformations. The model is formulated as:
\begin{equation}\label{GNN-HAR}
    \begin{split}
        \mathbf{H}^{(0)} =&\; (\mathbf{v}_d, \mathbf{v}_w, \mathbf{v}_m),\\
       \mathbf{H}^{(1)} =&\;  \text{ReLU}(\mathbf{D}^{-\frac{1}{2}} \mathbf{W} \mathbf{D}^{-\frac{1}{2}}\mathbf{H}^{(0)}\mathbf{W}^{(0)}),\\
        & \ldots,\\
        \mathbf{H}^{(L+1)} =&\; \text{ReLU}(\mathbf{D}^{-\frac{1}{2}} \mathbf{W} \mathbf{D}^{-\frac{1}{2}}\mathbf{H}^{(l)}\mathbf{W}^{(L)}),\\
        \widehat{\mathbf{v}}_t =&\; \boldsymbol{\alpha} + \beta_d \mathbf{v}_d + \beta_w \mathbf{v}_w + \beta_m \mathbf{v}_m + \boldsymbol{\gamma} \mathbf{H}^{(L+1)} +\boldsymbol{\epsilon}_t.
    \end{split}
\end{equation}
Here, $\{\mathbf{W}^{(l)}\}_{l=0}^L$ are trainable GNN parameters for the linear transformation, and ReLU$(\cdot)$ function transforms the aggregated volatility information nonlinearly. $\boldsymbol{\gamma}$ also contains trainable parameters to measure the overall influence of related stock market indices on the forecast. The GNN-HAR model uses spatial GNNs to capture the nonlinear relationships between stock market indices and generates more accurate RV forecasts compared to the HAR model \citep{Zhang2025}.

There are several other HAR model extensions, such as the VHAR model \citep{Bubak2011} and the HAR-KS model \citep{Liang2020}, which expand the linear combination of the HAR framework to incorporate historical volatility patterns of all entities in the system when generating RV forecasts for a target entity. However, these approaches are less powerful because they provide limited information about the structure of the volatility spillover network, and oversimplify the volatility spillover effect by assuming it is linear.

\section{Methodology and Implementation}
\label{Methodology and Implementation}
In this section, the GSE is defined and analyzed for volatility spillover networks constructed using different methods. Building on the insights from the GSE analysis, we then discuss the motivation, necessity and detailed implementation of the proposed GSP-HAR framework.

Denote the RV time series as $\{\mathbf{v}_t \in \mathbb{R}^N \}_{t = 1}^T$, which consists of the RV time series of $N$ different stock market indices on $T$ common trading days. Specifically, each $\mathbf{v}_t = (v_{1,t}, v_{2,t}, \dots, v_{N,t})^\top \in \mathbb{R}^N$ represents the RV observations of $N$ markets at trading day $t$.

\subsection{Volatility Spillover Network and Its GSE}
\label{Volatility Spillover Network and Its GSE}
Unlike traffic networks \citep{Li2018} where linkages correspond to tangible infrastructure, such as roads between cities, financial networks of stock market indices lack observable physical connections. Although the geographical distance between markets can be used to define volatility linkages, inferring the volatility spillover network of global stock market indices from data offers a more informative representation of inter-market volatility relationships. Here, three common data-driven methods for constructing volatility networks are discussed. These methods take the multi-asset RV time series $\{\mathbf{v}_t \in \mathbb{R}^N \}_{t = 1}^T$ as input, and output the weight matrix $\mathbf{W}$ of the volatility spillover network. The use of RV time series data, as opposed to return data, to construct the volatility spillover network is motivated by the argument that the volatility shows a higher degree of serial dependence than returns, particularly when measured at relatively high frequencies \citep{Diebold2014}. A brief summary of the four weight matrices produced by the three different volatility network construction methods is provided below, whereas their detailed calculation processes are described in Appendix \ref{Volatility Spillover Network Formulation}.
\begin{itemize}
    \item The Pearson correlation matrix method: the symmetric weight matrix is produced based on the Pearson correlation matrix according to Appendix \ref{The Pearson Correlation Matrix formulation}: $\mathbf{W} = \mathbf{W}^{P}$.
    \item The GLASSO precision matrix method: the symmetric weight matrix is produced based on the GLASSO precision matrix according to Appendix \ref{The GLASSO Precision Matrix formulation}: $\mathbf{W} = \mathbf{W}^{GL}$.
    \item The DY-symmetrization method: the symmetric weight matrix is produced based on the DY framework according to Appendix \ref{The DY Framework formulation}: $\mathbf{W} = \frac{1}{2} (\mathbf{W}^{DY} + (\mathbf{W}^{DY})^\top)$.
    \item The DY-magnetic-Laplacian method: the asymmetric weight matrix is produced based on the DY framework according to Appendix \ref{The DY Framework formulation}: $\mathbf{W} = \mathbf{W}^{DY}$. The magnetic Laplacian refers to the corresponding Laplacian matrix construction method for the asymmetric weight matrix.
\end{itemize}

As discussed in Section \ref{Graph Theory Fundamentals}, in order to calculate the non-negative GSE based on the normalized Laplacian matrix $\mathbf{L}$ (Equation \eqref{graph_signal_energy}), the weight matrix $\mathbf{W}$ of the volatility spillover network must be symmetric and can only contains non-negative elements, and the degree matrix $\mathbf{D}$ needs to be invertible. Suppose these conditions are always satisfied for the applied volatility spillover network construction methods. A rolling-window procedure is employed where the half window length, denoted as $\tau$, is set to approximately match the number of half-year common trading days across the selected stock market indices. For a given center date $t_0 \ (t_0 > \tau)$, the window extracts a one-year segment of RV data $\{\mathbf{v}_t \in \mathbb{R}^N\}_{t = t_0 - \tau}^{t_0 + \tau}$. It is the input to each graph construction method to generate the weight matrix $\mathbf{W}_{t_0}$ and, subsequently, the normalized graph Laplacian matrix $\mathbf{L}_{t_0}$. In addition, the yearly average vector RV, $\overline{\mathbf{v}}_{t_0} = (\overline{v}_{t_0,1}, \overline{v}_{t_0,2}, \dots, \overline{v}_{t_0,N})^\top$, is computed for each market within the window. Thus, the average GSE over the window centered at $t_0$ of the volatility spillover network can be calculated as:
\begin{equation}\label{VSP_signal_energy}
    E(\overline{\mathbf{v}}_{t_0}) = \overline{\mathbf{v}}_{t_0}^\top \mathbf{L}_{t_0} \overline{\mathbf{v}}_{t_0},
\end{equation}
which can be regarded as a function of $t_0$. For comparison purposes, the GSE $E(\overline{\mathbf{v}}_{t_0})$ of each graph construction method is normalized to the interval $[0,1]$ by dividing its all-time maximum value.

As ensuring the invertibility of the degree matrix is straightforward, the following discussion focuses on the conditions that the weight matrix must satisfy to produce a non-negative GSE. The Pearson correlation matrix method, the GLASSO precision matrix method and the DY-symmetrization method yield symmetric weight matrices that contain non-negative values. However, the DY-magnetic-Laplacian method produces an asymmetric weight matrix. Thus, additional operations are needed in this method to ensure a valid normalized Laplacian matrix for the corresponding GSE calculation. The normalized magnetic Laplacian $\mathbf{L}_m^{(q)} \in \mathbb{C}^{N \times N}$ is calculated based on the asymmetric $\mathbf{W}^{DY}$ to capture the directional information \citep{Zhang2021}. The formulation process is detailed in Section \ref{GSP Based on the Magnetic Laplacian}. The GSE value can also be measured for the magnetic Laplacian $\mathbf{L}_m^{(q)}$ through Equation \eqref{VSP_signal_energy} with $\mathbf{L}_{t_0} = \mathbf{L}_{m,t_0}^{(q)}$.

\begin{theorem}\label{Hermitian Magnetic Laplacian theorem}
For any graph $\mathcal{G}$, the normalized magnetic Laplacian $\mathbf{L}_m^{(q)} \in \mathbb{C}^{N \times N}$ is Hermitian, that is, $\mathbf{L}_m^{(q)} = (\mathbf{L}_m^{(q)})^*$.
\end{theorem}

\begin{theorem}\label{Real, Non-negative GSE Based on Hermitian Laplacian theorem}
The GSE based on the normalized magnetic Laplacian, calculated as $E(\mathbf{x}) = \mathbf{x}^\top \mathbf{L}_m^{(q)} \mathbf{x}$, is a real and non-negative scalar for graph signal $\mathbf{x} \in \mathbb{R}^N$.
\end{theorem}

Thus, the GSE of the DY–magnetic Laplacian method can be measured consistently with the other construction methods, which ensures comparability, and it remains a positive real number. Relevant proofs are given in Appendix \ref{Hermitian Magnetic Laplacian} and Appendix \ref{Real, Non-negative GSE Based on Hermitian Laplacian}.

The depiction of the rolling-window RV GSE $E(\overline{\mathbf{v}}_{t_0})$, calculated as Equation \eqref{VSP_signal_energy} for each graph construction method, is displayed in Appendix \ref{RV GSE Plots}. During turbulent market periods (e.g., financial crises), the cross-market volatility linkages become stronger \citep{Diebold2012, Bensaida2018}. The volatility spikes tend to be synchronous across markets, but with different magnitudes due to the distinct nature across markets. This is expected to lead to larger differences within RV graph signals $\overline{\mathbf{v}}_{t_0}$, which causes higher GSE levels. On the other hand, during calm market conditions, cross-market volatility interrelations are weaker \citep{Diebold2012, Bensaida2018}, and volatility levels are generally low and change slowly for various markets. This results in a smoother RV signal and, thus, lower RV GSE levels. It is intuitive that the RV GSE of the DY-magnetic-Laplacian volatility network best reflects the idea described above, and can be regarded as the most informative graph about the volatility interrelationships because it successfully identifies the significant financial crises and stable periods. 

In addition, by producing an asymmetric weight matrix of the volatility spillover network, the DY-magnetic-Laplacian method can capture the directional transmission of volatility. Influential stock markets tend to transmit more volatility to impressionable stock markets, whereas impressionable stock markets have less influence on influential stock markets. Thus, it is chosen to construct the volatility spillover network for the proposed RV forecasting model. In Section \ref{sec:Different_L}, we provide further empirical support for the choice.

\subsection{GSP Based on the Magnetic Laplacian}
\label{GSP Based on the Magnetic Laplacian}
Since the DY-magnetic-Laplacian construction method produces the most informative volatility spillover graph, the next step is to harness its structure to enhance the accuracy of RV forecasts. According to \citep{Zhang2021}, while spatial GNNs can be readily adapted to the directed volatility spillover graph due to their message-passing mechanism, they also have  certain limitations. Relying on the neighborhood information aggregation and transformation, spatial methods typically aggregate information from outgoing neighbors of a node. However, they place insufficient emphasis on the information from incoming neighbors. Thus, they may lose important volatility spillover information, which limits their ability to capture the volatility interrelationships. In contrast, the magnetic Laplacian is specifically designed to account for both incoming and outgoing connections, which allows it to preserve directional information and better capture the full structure of volatility spillovers \citep{Zhang2021}.

Suppose a directed graph $\mathcal{G}$ has $N$ nodes, which means $\mathbf{W} \in \mathbb{R}^{N \times N}$ and $\mathbf{W} \neq \mathbf{W}^\top$. In order to define the normalized magnetic Laplacian $\mathbf{L}_m^{(q)}$, the symmetric variant of weight matrix $\mathbf{W}^s$ and the corresponding degree matrix $\mathbf{D}^s$ are firstly defined in Equation \eqref{A^s} and Equation \eqref{D^s}, respectively,
\begin{equation}\label{A^s}
    \mathbf{W}^s = \frac{1}{2} (\mathbf{W} + \mathbf{W}^\top);
\end{equation}
\begin{equation}\label{D^s}
    \mathbf{D}^s = \text{diag}(\sum_{j = 1}^{N} \mathbf{W}^s_{1j}, \ldots, \sum_{j = 1}^{N} \mathbf{W}^s_{Nj}).
\end{equation}
Besides, to capture the directional information, the phase matrix $\boldsymbol{\Theta}^{(q)}$ is constructed as follows,
\begin{equation}\label{Theta}
    \boldsymbol{\Theta}^{(q)} = 2 \pi q (\mathbf{W} - \mathbf{W}^\top),
\end{equation}
where $q$ is a non-negative hyperparameter that determines how to handle the directionality of the graph. Use $\mathrm{i}$ to represent the imaginary unit. Based on $\mathbf{W}^s$ and $\boldsymbol{\Theta}^{(q)}$, the normalized magnetic Laplacians is formulated as, 
\begin{equation}\label{normalized magnetic Laplacian}
    \mathbf{L}_m^{(q)} = \mathbf{I} - ((\mathbf{D}^s)^{-\frac{1}{2}} \mathbf{W}^s (\mathbf{D}^s)^{-\frac{1}{2}}) \odot \exp(\mathrm{i} \boldsymbol{\Theta}^{(q)}),
\end{equation}
where $\exp(\cdot)$ is the exponential transformation applied to each element of the input matrix. $q$ determines how the directional information is perceived and processed. For example, if $q = 0$, $\mathbf{L}_m^{(0)} = \mathbf{I} - ((\mathbf{D}^s)^{-\frac{1}{2}} \mathbf{W}^s (\mathbf{D}^s)^{-\frac{1}{2}})$. The directionality of the graph is eliminated through symmetrizing the asymmetric weight matrix. For more discussions on how different $q$ values fit different graph motifs and reveal corresponding topological features, see \citep{Guo2017, Fanuel2018, Mohar2020, Zhou2020}. In this study, the DY-magnetic-Laplacian construction method requires $q>0$ to utilize the direction of the volatility spillover effect.

According to Theorem \ref{Hermitian Magnetic Laplacian theorem}, the normalized magnetic Laplacian matrix $\mathbf{L}_m^{(q)}$ is Hermitian. Besides, it is positive semi-definite \citep{Zhang2021}. The spectral theorem indicates that its eigendecomposition yields an orthonormal basis of complex eigenvectors with real and non-negative eigenvalues, presented as follows,
\begin{equation}\label{normalized magnetic Laplacian eigendecomposition}
    \mathbf{L}_m^{(q)} = \mathbf{U}_m \boldsymbol{\Lambda}_m \mathbf{U}_m^{\dagger},
\end{equation}
where the two complex matrices $\mathbf{U}_m$ and $\mathbf{U}_m^{\dagger}$ are unitary inverses of each other, i.e., $\mathbf{U}_m \mathbf{U}_m^{\dagger} = \mathbf{U}_m^{\dagger} \mathbf{U}_m = \mathbf{I}$, and the eigenvalues in $\boldsymbol{\Lambda}_m$ are real, non-negative numbers. The columns of $\mathbf{U}_m$ are eigenvectors of $\mathbf{L}_m^{(q)}$, and they form the orthonormal graph Fourier basis. Meanwhile, $\mathbf{U}_m^{\dagger}$ is the conjugate transpose of $\mathbf{U}_m$, and the rows of $\mathbf{U}_m^{\dagger}$ constitute another orthonormal basis dual to the graph Fourier basis in $\mathbb{C}^N$. Elements in $\boldsymbol{\Lambda}_m$ represent the frequency of the corresponding graph Fourier modes.

Since the DY-magnetic-Laplacian method is most informative about the inter-market volatility relationships and market volatility conditions, the orthonormal graph Fourier basis of the corresponding normalized magnetic Laplacian matrix is expected to construct a space where the representation of RV data becomes highly effective in capturing the volatility dynamics. To obtain such RV representation in the graph spectral domain, the Graph Fourier Transformation (GFT) and its inverse (IGFT) for the RV signal $\mathbf{v}_t \in \mathbb{R}^N$ $(t = 1, \ldots, T)$ based on the normalized magnetic Laplacian are, respectively, defined as follows,
\begin{align}
\widetilde{\mathbf{v}}_t &= \mathbf{U}_m^{\dagger} \mathbf{v}_t;  \label{GFT magnet}\\ 
\mathbf{v}_t &= \mathbf{U}_m \widetilde{\mathbf{v}}_t. \label{IGFT magnet}
\end{align}

\subsection{Model Formulation}
\begin{figure}[H]
    \centering
    \includegraphics[width=15cm]{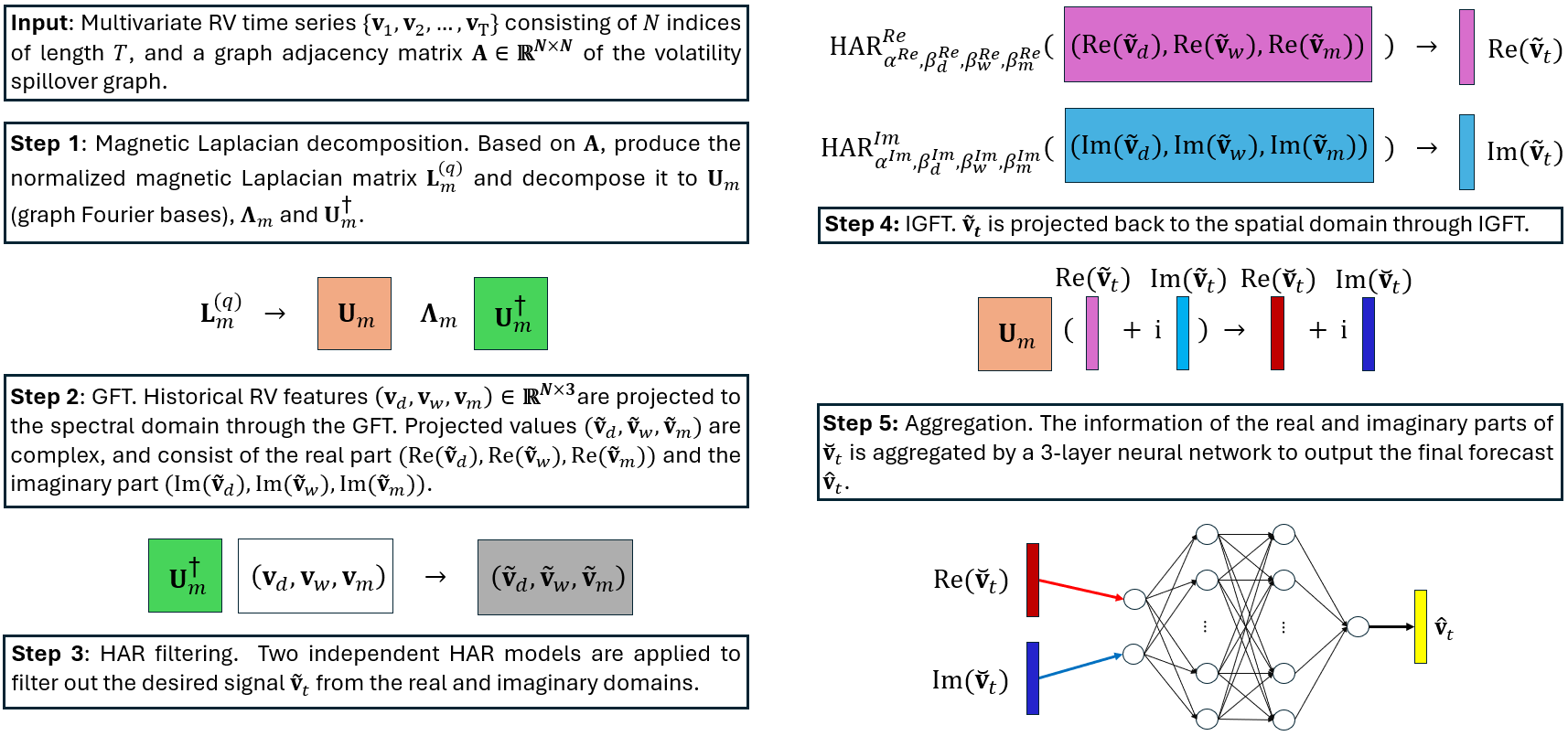}
    \caption{GSP-HAR model architecture}
    \label{GSP-HAR_model_architecture}
\end{figure}

In this section, we introduce the proposed GSP-HAR model, which combines graph signal processing techniques and the HAR framework to capture both temporal and cross-market volatility dynamics. The overall workflow and key components of the model are illustrated in Figure \ref{GSP-HAR_model_architecture}, providing a structured overview of how volatility spillover information is incorporated into the forecasting process.

Now we present the detailed step-by-step development of the proposed framework. In the first step, the weight matrix of the volatility spillover graph is produced under the DY framework and is denoted as $\mathbf{W}^{DY}$. The formulation process is illustrated in Appendix \ref{The DY Framework formulation}. Based on $\mathbf{W}^{DY}$, the corresponding normalized magnetic Laplacian matrix is calculated based on the process described from Equation \eqref{A^s} to Equation \eqref{normalized magnetic Laplacian} with $\mathbf{W} = \mathbf{W}^{DY}$. To effectively capture the directional volatility spillover effect between global stock markets, the hyperparameter $q$ of the magnetic Laplacian is positive, and its value is determined through the hyperparameter tuning process. Since $q > 0$, the eigendecomposition of the magnetic Laplacian $\mathbf{L}_m^{(q)}$ returns two complex matrices, $\mathbf{U}_m$ and $\mathbf{U}_m^{\dagger}$, and one eigenvalue matrix, $\boldsymbol{\Lambda}_m$, as shown in Equation \eqref{normalized magnetic Laplacian eigendecomposition}. Thus, the historical RV features $(\mathbf{v}_d, \mathbf{v}_w, \mathbf{v}_m) \in \mathbb{R}^{N \times 3}$ can be projected to the spectral domain through the GFT:
\begin{equation}\label{GSP-HAR-GFT}
    (\widetilde{\mathbf{v}}_d, \widetilde{\mathbf{v}}_w, \widetilde{\mathbf{v}}_m) = \mathbf{U}_m^{\dagger} (\mathbf{v}_d, \mathbf{v}_w, \mathbf{v}_m).
\end{equation}
Here, $(\widetilde{\mathbf{v}}_d, \widetilde{\mathbf{v}}_w, \widetilde{\mathbf{v}}_m)$ is a complex matrix. Its real part is denoted as $(\text{Re}(\widetilde{\mathbf{v}}_d), \text{Re}(\widetilde{\mathbf{v}}_w), \text{Re}(\widetilde{\mathbf{v}}_m))$, and its imaginary part is represented as $(\text{Im}(\widetilde{\mathbf{v}}_d), \text{Im}(\widetilde{\mathbf{v}}_w), \text{Im}(\widetilde{\mathbf{v}}_m))$. The spectral RV representation embeds the volatility interrelationships, which are encoded in the informative volatility spillover network based on the DY-magnetic-Laplacian method, into the historical RV pattern via the GFT process.

In the spectral domain, the GSP-HAR model leverages the HAR model as a linear data-driven filter within different parts of the basis. In the spectral domain, the HAR model filter takes the spectral RV representations as input signals and outputs new signals. The filtering processes for the real part and the imaginary part are independent, and their coefficients are drawn from two HAR filters so that the model can be more expressive. The filtering process of the two parts is formulated as:
\begin{equation} \label{real and imag HAR fixed}
    \begin{split}
        \text{Re}(\widetilde{\mathbf{v}}_t) =&\; \alpha^{Re}  + \beta_{d}^{Re} \text{Re}(\widetilde{\mathbf{v}}_d) + \beta_{w}^{Re} \text{Re}(\widetilde{\mathbf{v}}_w) + \beta_{m}^{Re} \text{Re}(\widetilde{\mathbf{v}}_m);\\
        \text{Im}(\widetilde{\mathbf{v}}_t) =&\; \alpha^{Im}  + \beta_{d}^{Im} \text{Im}(\widetilde{\mathbf{v}}_d) + \beta_{w}^{Im} \text{Im}(\widetilde{\mathbf{v}}_w) + \beta_{m}^{Im} \text{Im}(\widetilde{\mathbf{v}}_m).
    \end{split}
\end{equation}
The new signal $\widetilde{\mathbf{v}}_t \in \mathbb{C}^N$ and $\widetilde{\mathbf{v}}_t = \text{Re}(\widetilde{\mathbf{v}}_t) + \mathrm{i} \text{Im}(\widetilde{\mathbf{v}}_t)$. The two sets, $\{\alpha^{Re}, \beta_{d}^{Re}, \beta_{w}^{Re}, \beta_{m}^{Re}\}$ and $\{\alpha^{Im}, \beta_{d}^{Im}, \beta_{w}^{Im}, \beta_{m}^{Im}\}$, represent independent collections of HAR filter coefficients, all of which are real-valued. The new signal is projected back to the spatial domain through the IGFT:
\begin{equation}\label{fixeC-GSP-HAR-IGFT}
    \breve{\mathbf{v}}_t = \mathbf{U}_m \widetilde{\mathbf{v}}_t,
\end{equation}
where $\breve{\mathbf{v}}_t \in \mathbb{C}^N$ and $\breve{\mathbf{v}}_t = \text{Re}(\breve{\mathbf{v}}_t) + \mathrm{i} \text{Im}(\breve{\mathbf{v}}_t)$. In the spatial domain, the information of the real and imaginary parts of $\breve{\mathbf{v}}_t$ is aggregated by a $3$ layer shallow neural network $NN(\cdot)$ to output the final forecast $\widehat{\mathbf{v}}_t  \in \mathbb{R}^N$: 
\begin{equation}\label{fixeC-GSP-HAR-NN}
    \widehat{\mathbf{v}}_t = NN(\text{Re}(\breve{\mathbf{v}}_t), \text{Im}(\breve{\mathbf{v}}_t)).
\end{equation}
The neural network can merge the volatility information in the real and imaginary parts in a nonlinear way, which effectively enhances the expressive power of the proposed GSP-HAR model. As noted by \citet{Zhang2025}, such nonlinearity is crucial for accurately capturing volatility spillover effects and for improving the precision of RV forecasts.

\section{Experiments}
\label{sec: Experiments}
Through the intuitive GSE analysis discussed in previous sections, the DY-magnetic-Laplacian construction method is selected to construct the volatility spillover network due to its informativeness about market conditions. GSP techniques are subsequently leveraged to build a forecasting model specifically tailored to the selected graph construction method. To support this design, experiments are conducted using RV time series data from $24$ stock market indices. The main experiments assess the performance of the proposed GSP-HAR model by comparing it with baseline models to demonstrate its validity and effectiveness. Additional experiments further confirm that the model achieves superior forecasting performance when paired with the DY-magnetic-Laplacian construction method, compared to alternative approaches.

Moreover, as shown in Figure \ref{graph_energy_DY_magnet}, the GSE profile of the DY-magnetic-Laplacian method exhibits the potential to distinguish different financial market conditions. Building on this observation, an intuitive analysis is conducted to assess whether the GSP-HAR model, which leverages the informative DY-magnetic-Laplacian construction method, can provide early warnings of market turbulence.

\subsection{Model Evaluation Tools}
Various evaluation criteria are applied to evaluate the effectiveness of the proposed model. The evaluation process is conducted on an index-wise basis, as different markets operate under different economic conditions and investor behaviors. Especially, the accuracy of out-of-sample RV forecasts is measured through the commonly used Mean Squared Error (MSE) and Mean Absolute Error (MAE) criteria. Their calculation detail are shown in Equation \eqref{MSE_MAE_loss_cal} in Section \ref{sec:Main_Experiment_Results}. Both evaluation criteria are employed to ensure that the models perform consistently, thereby providing a more robust assessment of their forecast accuracy. In addition, the Model Confidence Set (MCS, \citealt{Hansen2011}) test is applied to evaluate all tested models to determine a subset of models that have statistically superior forecasting accuracy (based on MSE or MAE) for each stock market index. Similar to the input settings of the HAR models, the forecasting windows are set as $h = 1, 5, 22$, corresponding to the short-term (daily), mid-term (weekly) and long-term (monthly) forecasting. The software and hardware configurations are:
\begin{itemize}
    \item Software: Python 3.10.14; NumPy 1.26.4; Statsmodels 0.14.2; PyTorch 2.2.2+cu121.
    \item Hardware: Operating system: Windows 11; CPU: 13th Gen Intel(R) Core(TM) i9-13900HX; GPU: NVIDIA GeForce RTX 4080 Laptop GPU.
\end{itemize}

\subsection{Benchmark Datasets and Baseline models}
The benchmark dataset, which is downloaded from Oxford-Man Institute's realized library, consists of the RV data of $24$ stock market indices with around $3500$ common trading days from May 2002 to June 2022. Among them, some indices are commonly used indices in previous RV forecast research (e.g., \citep{Liang2020}). The daily RV data is produced based on the $5$-minute intraday high frequency returns. Given that RV forecast models are trained on RV observations on common trading days of the included stock market indices, the selection of these $24$ indices is intentionally focused on markets with more overlapping trading schedules. This approach ensures that the RV time series can be more continuous with more synchronized trading days across different markets, thus improving the data usage efficiency and enhancing the reliability and comparability of the RV forecasts produced by different models. The first $70\%$ of data (from May 2002 to Feb 2016) are the in-sample data, whereas the last $30\%$ of data (from March 2016 to June 2022) are the out-of-sample data. Besides, the RV is scaled up by multiplying $100$. A descriptive analysis of the scaled RV data for each stock market index is reported in Table \ref{rv_stats} in Appendix \ref{RV Time Series Data Statistics}. 

Figure \ref{Net_VSP_network} shows the volatility spillover network for $7$ selected stock market indices for demonstration and clarity. The nodes are the abbreviation of the selected indices, and the width of edges is proportional to the scale of the volatility spillover effect calculated under the DY framework. The arrows of the edges indicate the direction of the volatility spillover effect.
\begin{figure}[H]
    \centering
    \includegraphics[width=10cm]{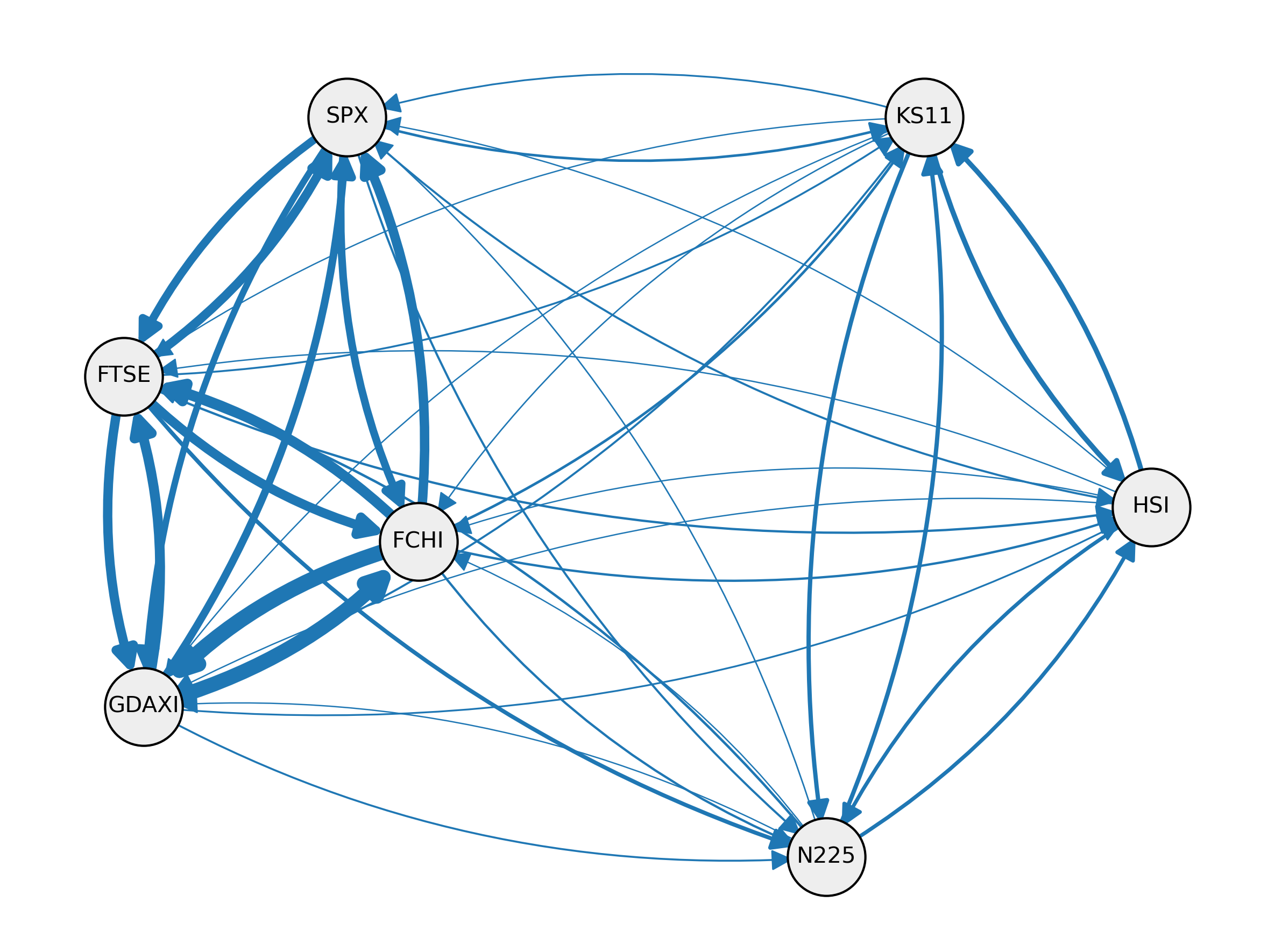}
    \caption{A schematic demonstration of the volatility spillover network}
    \label{Net_VSP_network}
\end{figure}

The baseline models include the HAR, VHAR, HAR-KS and GNN-HAR models. Although the exact construction of some baseline models is not available, they are formulated according to the corresponding research papers. For the GNN-HAR model, different values are tried for the hyperparameter $L$, and only the best performance is reported in each RV forecasting task.

\subsection{Experimental Results}\label{sec:Main_Experiment_Results}
The general index-wise performance of each model is measured by the MSE and MAE calculated as follows:
\begin{equation} \label{MSE_MAE_loss_cal}
\begin{split}
    \text{MSE}_i =&\; \frac{1}{T_{out}} \sum_{t=1}^{T_{out}}(\widehat{\mathbf{v}}_{i,t} - \mathbf{v}_{i,t})^2;\\
    \text{MAE}_i =&\; \frac{1}{T_{out}} \sum_{t=1}^{T_{out}}|\widehat{\mathbf{v}}_{i,t} - \mathbf{v}_{i,t}|,
\end{split}
\end{equation}
where $\text{MSE}_i$ and $\text{MAE}_i$ is the MSE and MAE of the $i_{\text{th}}$ stock market index, respectively. $T_{out}$ denotes the duration of the out-of-sample dataset. $\widehat{\mathbf{v}}_{i,t}$ refers to the RV forecasting value of the $i_{\text{th}}$ index at time $t$, and $\mathbf{v}_{i,t}$ is the corresponding true RV value. The index-wise MSE results, MAE results and the MCS test results for different models across various forecasting horizons ($h = 1, 5, 22$) are presented from Table \ref{tab:mse_mae_model_h1} to Table \ref{tab:mse_mae_model_h22}.

\begin{table}[H]
\centering
\footnotesize
\setlength{\tabcolsep}{3pt}
\renewcommand{\arraystretch}{0.95}
\begin{tabular}{l|ccccc|ccccc}
\toprule
         & \multicolumn{5}{c|}{MSE ($h=1$)}                     & \multicolumn{5}{c}{MAE ($h=1$)} \\
\cmidrule(lr){2-6}\cmidrule(lr){7-11}
         & HAR              & VHAR             & HAR-KS           & GNN-HAR           & GSP-HAR           & HAR              & VHAR             & HAR-KS           & GNN-HAR           & GSP-HAR           \\
\midrule
AEX      & \cg{0.091}       & \cg{0.094}       & \cg{0.089}       & \cg{0.091}       & \cb{\cg{0.089}} & 0.184            & 0.199            & 0.187            & \cg{0.180}       & \cb{\cg{0.179}}  \\
AORD     & 0.111            & 0.131            & \cg{0.109}       & \cg{0.110}       & \cb{\cg{0.108}} & 0.184            & 0.266            & 0.217            & \cg{0.180}       & \cb{\cg{0.178}}  \\
BFX      & \cg{0.086}       & 0.092            & \cg{0.084}       & \cg{0.085}       & \cb{\cg{0.084}} & 0.180            & 0.191            & \cg{0.179}       & \cb{\cg{0.176}}  & \cg{0.177}       \\
BSESN    & 0.125            & 0.118            & \cb{\cg{0.107}}  & 0.124            & \cg{0.117}      & 0.170            & 0.207            & 0.182            & \cg{0.166}       & \cb{\cg{0.164}}  \\
BVSP     & \cg{0.122}       & 0.164            & 0.134            & \cg{0.126}       & \cb{\cg{0.121}} & \cg{0.204}       & 0.256            & 0.220            & \cg{0.203}       & \cb{\cg{0.203}}  \\
DJI      & 0.112            & 0.140            & 0.116            & \cb{\cg{0.111}}  & 0.112           & 0.194            & 0.257            & 0.212            & \cb{\cg{0.189}}  & \cg{0.190}       \\
FCHI     & \cg{0.086}       & \cg{0.097}       & \cg{0.089}       & \cg{0.086}       & \cb{\cg{0.083}} & 0.192            & 0.199            & \cg{0.190}       & \cb{\cg{0.189}}  & \cg{0.189}       \\
FTSE     & 0.205            & 0.194            & \cb{\cg{0.185}}  & 0.205            & 0.201           & 0.233            & \cg{0.224}       & \cb{\cg{0.219}}  & \cg{0.230}       & \cg{0.229}       \\
GDAXI    & \cg{0.072}       & 0.082            & 0.075            & \cg{0.073}       & \cb{\cg{0.068}} & \cg{0.178}       & 0.192            & \cg{0.183}       & \cg{0.176}       & \cb{\cg{0.174}}  \\
GSPTSE   & 0.062            & 0.134            & 0.086            & \cg{0.061}       & \cb{\cg{0.060}} & 0.135            & 0.300            & 0.207            & \cb{\cg{0.129}}  & \cg{0.129}       \\
HSI      & \cg{0.072}       & 0.114            & 0.086            & \cg{0.072}       & \cb{\cg{0.070}} & 0.157            & 0.206            & 0.176            & \cg{0.154}       & \cb{\cg{0.154}}  \\
IBEX     & \cg{0.128}       & 0.161            & 0.143            & \cg{0.129}       & \cb{\cg{0.127}} & \cg{0.200}       & 0.219            & \cg{0.202}       & \cb{\cg{0.198}}  & \cg{0.199}       \\
IXIC     & \cg{0.115}       & 0.141            & 0.119            & \cg{0.115}       & \cb{\cg{0.114}} & 0.219            & 0.260            & 0.226            & \cg{0.215}       & \cb{\cg{0.215}}  \\
KS11     & \cg{0.062}       & 0.091            & \cg{0.071}       & \cb{\cg{0.062}}  & \cg{0.062}      & 0.144            & 0.215            & 0.180            & \cg{0.140}       & \cb{\cg{0.139}}  \\
KSE      & \cg{0.102}       & 0.154            & 0.121            & \cb{\cg{0.102}}  & \cg{0.102}      & 0.199            & 0.287            & 0.245            & \cg{0.196}       & \cb{\cg{0.194}}  \\
MXX      & \cg{0.081}       & 0.114            & 0.094            & \cb{\cg{0.081}}  & \cg{0.082}      & 0.177            & 0.204            & 0.187            & \cb{\cg{0.174}}  & \cg{0.174}       \\
N225     & 0.117            & 0.121            & \cb{\cg{0.110}}  & \cg{0.117}       & \cg{0.116}      & 0.190            & 0.230            & 0.203            & \cg{0.185}       & \cb{\cg{0.183}}  \\
NSEI     & 0.131            & 0.125            & \cb{\cg{0.112}}  & 0.130            & \cg{0.126}      & 0.171            & 0.216            & 0.187            & \cg{0.167}       & \cb{\cg{0.165}}  \\
OSEAX    & 0.348            & 0.325            & \cb{\cg{0.311}}  & 0.348            & 0.336           & 0.267            & \cg{0.246}       & \cb{\cg{0.244}}  & 0.263            & 0.264            \\
RUT      & \cg{0.101}       & 0.107            & \cb{\cg{0.093}}  & \cg{0.101}       & \cg{0.099}      & 0.206            & 0.226            & \cg{0.204}       & \cb{\cg{0.203}}  & \cg{0.204}       \\
SPX      & \cg{0.113}       & 0.145            & 0.120            & \cg{0.116}       & \cb{\cg{0.110}} & \cg{0.200}       & 0.274            & 0.226            & \cg{0.197}       & \cb{\cg{0.195}}  \\
SSEC     & 0.063            & 0.163            & 0.110            & \cb{\cg{0.062}}  & 0.064           & 0.172            & 0.263            & 0.220            & \cb{\cg{0.169}}  & \cg{0.171}       \\
SSMI     & 0.089            & 0.119            & 0.100            & \cg{0.088}       & \cb{\cg{0.084}} & 0.138            & 0.211            & 0.172            & \cg{0.134}       & \cb{\cg{0.133}}  \\
STOXX50E & \cg{0.166}       & 0.184            & \cg{0.165}       & \cg{0.166}       & \cb{\cg{0.164}} & \cg{0.227}       & \cg{0.229}       & \cb{\cg{0.219}}  & \cg{0.224}       & \cg{0.225}       \\
\bottomrule
\end{tabular}
\caption{MSE and MAE results comparisons across different models at the short-term forecasting horizon ($h=1$). Cells highlighted in gray are models selected by the MCS test at the $25\%$ significance level; numbers in blue indicate the best-performing model per index.}
\label{tab:mse_mae_model_h1}
\end{table}

\begin{table}[H]
\centering
\footnotesize
\setlength{\tabcolsep}{3pt}
\renewcommand{\arraystretch}{0.95}
\begin{tabular}{l|ccccc|ccccc}
\toprule
         & \multicolumn{5}{c|}{MSE ($h=5$)}                     & \multicolumn{5}{c}{MAE ($h=5$)} \\
\cmidrule(lr){2-6}\cmidrule(lr){7-11}
         & HAR              & VHAR             & HAR-KS           & GNN-HAR           & GSP-HAR           & HAR              & VHAR             & HAR-KS           & GNN-HAR           & GSP-HAR           \\
\midrule
AEX      & \cg{0.163}       & 0.204            & \cg{0.161}       & \cg{0.159}       & \cb{\cg{0.156}}  & 0.242            & 0.325            & 0.252            & 0.241            & \cb{\cg{0.233}}\\
AORD     & \cg{0.166}       & \cg{0.162}       & \cg{0.161}       & \cg{0.161}       & \cb{\cg{0.157}}  & 0.210            & 0.228            & \cg{0.203}       & \cb{\cg{0.201}}           & 0.225          \\
BFX      & \cg{0.142}       & 0.164            & \cb{\cg{0.138}}  & \cg{0.141}       & \cg{0.139}       & \cg{0.225}       & 0.277            & \cg{0.225}       & \cg{0.223}                & \cb{\cg{0.221}}\\
BSESN    & 0.203            & 0.245            & 0.213            & 0.195            & \cb{\cg{0.185}}  & 0.236            & 0.313            & 0.263            & 0.217            & \cb{\cg{0.208}}\\
BVSP     & \cg{0.209}       & 0.247            & \cb{\cg{0.199}}  & \cg{0.206}       & \cg{0.203}       & 0.274            & 0.369            & 0.273            & \cg{0.255}       & \cb{\cg{0.248}}\\
DJI      & \cg{0.212}       & 0.234            & \cg{0.206}       & \cg{0.207}       & \cb{\cg{0.205}}  & \cg{0.266}       & 0.321            & \cg{0.267}       & \cg{0.270}       & \cb{\cg{0.266}}\\
FCHI     & \cg{0.162}       & 0.227            & \cg{0.161}       & \cg{0.159}       & \cb{\cg{0.156}}  & 0.253            & 0.368            & 0.263            & \cg{0.248}       & \cb{\cg{0.243}}\\
FTSE     & \cg{0.271}       & \cg{0.279}       & \cb{\cg{0.263}}  & \cg{0.268}       & \cg{0.264}       & \cg{0.279}       & 0.320            & \cg{0.274}       & \cg{0.278}            & \cb{\cg{0.270}}\\
GDAXI    & \cg{0.137}       & 0.196            & \cg{0.134}       & \cg{0.132}       & \cb{\cg{0.128}}  & \cg{0.237}       & 0.343            & 0.250            & \cg{0.231}       & \cb{\cg{0.227}}\\
GSPTSE   & \cg{0.119}       & 0.135            & \cg{0.121}       & \cg{0.111}       & \cb{\cg{0.106}}  & \cb{\cg{0.181}}  & 0.231            & 0.196            & 0.197            & \cg{0.195}     \\
HSI      & \cb{\cg{0.100}}  & 0.111            & \cg{0.100}       & \cg{0.102}       & \cg{0.104}       & \cb{\cg{0.186}}  & 0.217            & \cg{0.189}       & \cg{0.193}       & \cg{0.189}     \\
IBEX     & 0.225            & 0.298            & 0.229            & 0.222            & \cb{\cg{0.213}}  & 0.265            & 0.398            & 0.287            & 0.253            & \cb{\cg{0.246}}\\
IXIC     & \cg{0.222}       & \cg{0.222}       & \cb{\cg{0.217}}  & \cg{0.218}       & \cg{0.217}       & \cg{0.294}       & 0.321            & \cb{\cg{0.290}}  & 0.298            & \cg{0.294}     \\
KS11     & \cg{0.101}       & \cg{0.092}       & \cb{\cg{0.091}}  & \cg{0.100}       & \cg{0.097}       & \cg{0.183}       & \cg{0.185}       & \cg{0.184}       & \cg{0.182}       & \cb{\cg{0.179}}\\
KSE      & \cg{0.138}       & 0.148            & \cg{0.134}       & \cg{0.139}       & \cb{\cg{0.134}}  & 0.237            & \cg{0.239}       & \cg{0.231}       & \cg{0.235}       & \cb{\cg{0.227}}\\
MXX      & \cb{\cg{0.095}}  & 0.111            & \cg{0.096}       & \cg{0.098}       & \cg{0.099}       & \cb{\cg{0.193}}  & 0.235            & \cg{0.194}       & 0.201            & \cg{0.197}     \\
N225     & \cg{0.165}       & 0.180            & \cb{\cg{0.161}}  & \cg{0.166}       & \cg{0.163}       & 0.250            & 0.272            & 0.250            & 0.244            & \cb{\cg{0.237}}\\
NSEI     & 0.209            & 0.269            & 0.227            & 0.199            & \cb{\cg{0.188}}  & 0.242            & 0.342            & 0.277            & 0.217            & \cb{\cg{0.207}}\\
OSEAX    & \cg{0.409}       & 0.457            & \cg{0.406}       & \cg{0.406}       & \cb{\cg{0.399}}  & 0.317            & 0.402            & 0.317            & 0.311            & \cb{\cg{0.299}}\\
RUT      & \cg{0.189}       & \cg{0.185}       & \cb{\cg{0.181}}  & \cg{0.187}       & \cg{0.185}       & \cg{0.264}       & 0.288            & \cg{0.261}       & \cb{\cg{0.261}}  & \cg{0.264}     \\
SPX      & \cg{0.222}       & 0.241            & \cg{0.218}       & \cg{0.218}       & \cb{\cg{0.217}}  & \cg{0.280}  & 0.342            & \cg{0.283}       & \cb{\cg{0.278}}       & \cg{0.285}     \\
SSEC     & 0.100            & 0.166            & 0.118            & \cb{\cg{0.094}}  & \cg{0.096}       & 0.240            & 0.309            & 0.271            & 0.215            & \cb{\cg{0.210}}\\
SSMI     & \cg{0.152}       & 0.181            & \cg{0.151}       & \cg{0.151}       & \cb{\cg{0.141}}  & \cg{0.180}       & 0.250            & 0.189            & \cg{0.179 }      & \cb{\cg{0.177}}\\
STOXX50E & \cg{0.262}       & 0.319            & \cg{0.259}       & \cg{0.257}       & \cb{\cg{0.251}}  & 0.295            & 0.404            & 0.310            & \cg{0.285}       & \cb{\cg{0.277}}\\
\bottomrule
\end{tabular}
\caption{MSE and MAE results comparisons across different models at the mid-term forecasting horizon ($h=5$). Cells highlighted in gray are models selected by the MCS test at the $25\%$ significance level; numbers in blue indicate the best-performing model per index.}
\label{tab:mse_mae_model_h5}
\end{table}

\begin{table}[H]
\centering
\footnotesize
\setlength{\tabcolsep}{3pt}
\renewcommand{\arraystretch}{0.95}
\begin{tabular}{l|ccccc|ccccc}
\toprule
         & \multicolumn{5}{c|}{MSE ($h=22$)}                     & \multicolumn{5}{c}{MAE ($h=22$)} \\
\cmidrule(lr){2-6}\cmidrule(lr){7-11}
         & HAR              & VHAR             & HAR-KS           & GNN-HAR           & GSP-HAR           & HAR              & VHAR             & HAR-KS           & GNN-HAR           & GSP-HAR           \\
\midrule
AEX      & \cg{0.240}     & 0.294          & 0.246          & \cg{0.237}     & \cb{\cg{0.230}} & 0.290          & 0.366          & 0.306            & \cg{0.279}     & \cb{\cg{0.266}} \\
AORD     & \cg{0.241}     & \cg{0.238}     & \cb{\cg{0.232}}& \cg{0.239}     & 0.252           & \cb{\cg{0.248}}& \cg{0.259}     & \cg{0.249}       & 0.276          & 0.310           \\
BFX      & \cg{0.199}     & \cg{0.222}     & \cg{0.198}     & \cb{\cg{0.198}}& \cg{0.202}      & \cg{0.256}     & 0.306          & 0.265            & \cb{\cg{0.256}}& \cg{0.260}      \\
BSESN    & \cg{0.287}     & 0.353          & \cg{0.281}     & \cg{0.275}     & \cb{\cg{0.264}} & 0.307          & 0.406          & 0.319            & \cb{\cg{0.260}}& \cg{0.268}      \\
BVSP     & \cg{0.314}     & 0.420          & \cg{0.311}     & \cg{0.336}     & \cb{\cg{0.304}} & 0.351          & 0.486          & 0.367            & 0.319          & \cb{\cg{0.291}} \\
DJI      & \cg{0.334}     & 0.374          & 0.348          & \cg{0.337}     & \cb{\cg{0.321}} & \cg{0.341}     & 0.398          & 0.365            & \cg{0.335}     & \cb{\cg{0.334}} \\
FCHI     & \cg{0.241}     & 0.344          & 0.258          & \cg{0.236}     & \cb{\cg{0.233}} & 0.302          & 0.451          & 0.341            & \cg{0.281}     & \cb{\cg{0.273}} \\
FTSE     & \cg{0.370}     & \cg{0.385}     & \cg{0.372}     & \cg{0.367}     & \cb{\cg{0.355}} & 0.329          & 0.366          & 0.338            & \cg{0.318}     & \cb{\cg{0.303}} \\
GDAXI    & 0.208          & 0.297          & 0.224          & \cg{0.198}     & \cb{\cg{0.196}} & 0.287          & 0.402          & 0.322            & \cb{\cg{0.260}}& \cg{0.270}      \\
GSPTSE   & \cg{0.194}     & 0.268          & 0.214          & \cg{0.192}     & \cb{\cg{0.185}} & \cb{\cg{0.234}}& 0.323          & 0.276            & 0.257          & 0.278           \\
HSI      & 0.119          & 0.160          & 0.127          & \cg{0.117}     & \cb{\cg{0.116}} & \cg{0.214}     & 0.272          & 0.233            & \cg{0.211}     & \cb{\cg{0.211}} \\
IBEX     & \cg{0.281}     & 0.398          & 0.303          & \cb{\cg{0.278}}& \cg{0.278}      & 0.303          & 0.484          & 0.360            & \cb{\cg{0.275}}& \cg{0.288}      \\
IXIC     & \cg{0.319}     & \cg{0.342}     & \cg{0.321}     & \cg{0.323}     & \cb{\cg{0.319}} & \cg{0.353}     & 0.390          & \cg{0.358}       & \cg{0.354}     & \cb{\cg{0.339}} \\
KS11     & \cg{0.132}     & 0.166          & \cb{\cg{0.128}}& \cg{0.131}     & \cg{0.130}      & \cg{0.221}     & \cg{0.256}     & \cb{\cg{0.217}}  & \cg{0.218}     & \cg{0.219}      \\
KSE      & \cg{0.181}     & 0.293          & \cg{0.181}     & \cg{0.182}     & \cb{\cg{0.179}} & \cg{0.281}     & 0.358          & \cb{\cg{0.267}}  & \cg{0.267}     & \cg{0.274}      \\
MXX      & 0.114          & 0.147          & 0.118          & 0.115          & \cb{\cg{0.112}} & \cg{0.214}     & 0.270          & \cg{0.218}       & \cg{0.218}     & \cb{\cg{0.213}} \\
N225     & \cg{0.205}     & 0.233          & \cg{0.212}     & \cg{0.204}     & \cb{\cg{0.203}} & 0.299          & 0.303          & 0.310            & \cb{\cg{0.271}}& \cg{0.282}      \\
NSEI     & 0.297          & 0.373          & 0.292          & 0.281          & \cb{\cg{0.265}} & 0.315          & 0.422          & 0.328            & \cb{\cg{0.264}}& \cg{0.268}      \\
OSEAX    & \cg{0.526}     & 0.623          & \cg{0.532}     & \cg{0.532}     & \cb{\cg{0.506}} & 0.376          & 0.492          & 0.393            & \cg{0.352}     & \cb{\cg{0.333}} \\
RUT      & \cg{0.268}     & \cg{0.286}     & \cg{0.268}     & \cg{0.271}     & \cb{\cg{0.266}} & \cg{0.310}     & 0.355          & 0.319            & 0.316          & \cb{\cg{0.305}} \\
SPX      & \cg{0.345}     & 0.409          & \cg{0.358}     & \cg{0.347}     & \cb{\cg{0.330}} & \cg{0.354}     & 0.433          & 0.379            & \cb{\cg{0.350}}& \cg{0.369}      \\
SSEC     & 0.146          & 0.290          & 0.190          & 0.117          & \cb{\cg{0.111}} & 0.313          & 0.424          & 0.362            & \cg{0.248}     & \cb{\cg{0.240}} \\
SSMI     & \cg{0.252}     & 0.289          & \cg{0.251}     & \cg{0.253}     & \cb{\cg{0.235}} & \cg{0.237}     & 0.297          & \cg{0.243}       & \cg{0.241}     & \cb{\cg{0.237}} \\
STOXX50E & \cg{0.358}     & 0.443          & \cg{0.372}     & \cg{0.354}     & \cb{\cg{0.346}} & 0.351          & 0.477          & 0.388            & \cg{0.319}     & \cb{\cg{0.315}} \\
\bottomrule
\end{tabular}
\caption{MSE and MAE results comparisons across different models at long-term forecasting horizon ($h=22$). Cells highlighted in gray are models selected by the MCS test at the $25\%$ significance level; numbers in blue indicate the best-performing model per index.}
\label{tab:mse_mae_model_h22}
\end{table}

According to the MSE and MAE results in the tables above, the GSP-HAR model improves the RV forecasting performance across all three forecast horizons. It consistently achieves the lowest MSE and MAE scores for more than half of the stock market indices in the short-term ($h = 1$), mid-term ($h = 5$) and long-term ($h = 22$) RV forecasting tasks. In addition, the MCS test is conducted to assess whether these improvements are statistically significant. Across different evaluation criteria and forecasting horizons, the GSP-HAR model remains in the MCS for more than $20$ stock market indices. On the other hand, other models appear in the MCS less frequently. These imply the consistent and significant superiority of the GSP-HAR model in different forecasting horizons. Compared to linear HAR models, such as HAR-KS and VHAR, the GSP-HAR is informative of the volatility interrelationships and is able to capture nonlinear volatility spillover effects, which is beneficial for accurate RV forecasting \citep{Zhang2025}. 

In contrast to the spatial GNN-HAR model, which faces challenges in capturing directional spillover information \citep{Zhang2021}, the spectral GSP-HAR model leverages GFT techniques based on the magnetic Laplacian to effectively model directional volatility spillovers. Besides, while adding additional GNN layers to the spatial GNN-HAR model yields little improvement in RV forecasting accuracy, consistent with the findings of \citep{Zhang2025}, the GSP-HAR model attains optimal performance by incorporating a global perspective on the volatility spillover effect. This contrast arises because multi-layer spatial GNNs aggregate information sequentially from local neighborhoods via multi-hop message passing, whereas GSP methods transform signals into the spectral domain, where the entire network structure is encoded and global interdependencies are effectively captured.

\subsection{GSP-HAR Forecasting Results with Different Laplacian Construction}\label{sec:Different_L}
Although the RV GSE analysis in previous sections indicates that the DY-magnetic-Laplacian method most effectively captures volatility interrelationships by reflecting market conditions, its superior performance in RV forecasting relative to other construction methods has yet to be empirically validated. In this section, the impact of different volatility network construction and the associated Laplacian formulation on GSP-HAR RV forecasting accuracy is analyzed. As shown in Appendix \ref{Volatility Spillover Network Formulation}, there are three widely-applied, data-driven volatility spillover network formulation methods of interest: the Pearson correlation matrix, the GLASSO precision matrix and the DY framework.

According to \citet{Zhang2021}, the magnetic Laplacian is a unified GSP framework which can be applied to both undirected and directed graphs by controlling the hyperparameter $q$. When $q = 0$, the volatility spillover is regarded as undirected, and the magnetic Laplacian matrix is identical to the common Laplacian matrix based on the symmetric weight matrix. On the other hand, the directional transmission of volatility can be processed when $q > 0$. Thus, the GSP-HAR model can handle all different volatility spillover network construction methods by choosing the appropriate $q$ value without changing the model architecture. The notations of GSP-HAR models with different graph inputs are summarized below:
\begin{itemize}
    \item `GSP-HAR-P': the weight matrix is symmetric: $\mathbf{W} = \mathbf{W}^{P}$. The hyperparameter $q$ in the GSP-HAR model is set to $0$: $q = 0$.
    \item `GSP-HAR-GL': the weight matrix is symmetric: $\mathbf{W} = \mathbf{W}^{GL}$. The hyperparameter $q$ in the GSP-HAR model is set to $0$: $q = 0$.
    \item `GSP-HAR-DY-sym': the weight matrix is symmetric: $\mathbf{W} = \frac{1}{2} (\mathbf{W}^{DY} + (\mathbf{W}^{DY})^\top)$. The hyperparameter $q$ in the GSP-HAR model is set to $0$: $q = 0$.
    \item `GSP-HAR-DY-asym': the weight matrix is asymmetric: $\mathbf{W} = \mathbf{W}^{DY}$. The hyperparameter $q$ in the GSP-HAR model is positive: $q > 0$. This is the same GSP-HAR model tested in Section \ref{sec:Main_Experiment_Results}.
\end{itemize}
The calculation process of $\mathbf{W}^{P}$, $\mathbf{W}^{GL}$ and $\mathbf{W}^{DY}$ are detailed in Appendix \ref{Volatility Spillover Network Formulation}. Similar to Section \ref{sec:Main_Experiment_Results}, the index-wise MSE results, MAE results and the MCS test results for different volatility network construction methods and different forecasting horizons ($h = 1, 5, 22$) are presented from Table \ref{tab:mse_mae_laplacian_h1} to Table \ref{tab:mse_mae_laplacian_h22}.

\begin{table}[H]
\centering
\footnotesize
\setlength{\tabcolsep}{3pt}
\renewcommand{\arraystretch}{0.95}
\begin{tabular}{l|cccc|cccc}
\toprule
         & \multicolumn{4}{c|}{MSE ($h=1$)}                     & \multicolumn{4}{c}{MAE ($h=1$)} \\
\cmidrule(lr){2-5}\cmidrule(lr){6-9}
GSP-HAR   & P               & GL              & DY-sym          & DY-asym         & P               & GL              & DY-sym          & DY-asym         \\
\midrule
AEX      & \cg{0.090}      & \cg{0.090}      & \cg{0.090}      & \cb{\cg{0.089}} & 0.180           & 0.180           & \cg{0.180}      & \cb{\cg{0.179}} \\
AORD     & \cg{0.109}      & \cg{0.109}      & \cg{0.109}      & \cb{\cg{0.108}} & 0.180           & 0.180           & 0.180           & \cb{\cg{0.178}} \\
BFX      & \cg{0.085}      & \cg{0.085}      & \cg{0.085}      & \cb{\cg{0.084}} & \cg{0.177}      & \cg{0.177}      & \cb{\cg{0.177}} & \cg{0.177}      \\
BSESN    & \cg{0.121}      & \cg{0.121}      & \cg{0.121}      & \cb{\cg{0.117}} & 0.166           & 0.166           & \cg{0.166}      & \cb{\cg{0.164}} \\
BVSP     & \cg{0.123}      & \cg{0.123}      & \cg{0.123}      & \cb{\cg{0.121}} & \cb{\cg{0.202}} & \cg{0.202}      & \cg{0.202}      & \cg{0.203}      \\
DJI      & \cg{0.111}      & \cb{\cg{0.110}} & \cg{0.111}      & 0.112           & \cg{0.190}      & \cg{0.191}      & \cg{0.190}      & \cb{\cg{0.190}} \\
FCHI     & \cg{0.086}      & \cg{0.086}      & \cg{0.086}      & \cb{\cg{0.083}} & 0.193           & \cg{0.191}      & \cg{0.191}      & \cb{\cg{0.189}} \\
FTSE     & \cg{0.200}      & \cg{0.201}      & \cb{\cg{0.200}} & \cg{0.201}      & \cg{0.230}      & \cg{0.230}      & \cg{0.230}      & \cb{\cg{0.229}} \\
GDAXI    & 0.071           & 0.071           & 0.071           & \cb{\cg{0.068}} & 0.176           & 0.176           & \cg{0.176}      & \cb{\cg{0.174}} \\
GSPTSE   & \cb{\cg{0.059}} & 0.059           & \cg{0.059}      & 0.060           & \cg{0.129}      & 0.130           & \cg{0.129}      & \cb{\cg{0.129}} \\
HSI      & \cg{0.072}      & \cg{0.072}      & \cg{0.072}      & \cb{\cg{0.070}} & \cg{0.155}      & \cg{0.155}      & \cg{0.155}      & \cb{\cg{0.154}} \\
IBEX     & \cg{0.129}      & \cg{0.129}      & \cg{0.129}      & \cb{\cg{0.127}} & \cg{0.199}      & \cb{\cg{0.199}} & 0.199           & \cg{0.199}      \\
IXIC     & \cg{0.116}      & \cg{0.116}      & \cg{0.116}      & \cb{\cg{0.114}} & 0.217           & 0.217           & 0.217           & \cb{\cg{0.215}} \\
KS11     & \cg{0.061}      & \cg{0.061}      & \cb{\cg{0.061}} & \cg{0.062}      & \cg{0.140}      & 0.140           & \cg{0.140}      & \cb{\cg{0.139}} \\
KSE      & \cg{0.101}      & \cb{\cg{0.101}} & \cg{0.101}      & \cg{0.102}      & 0.196           & 0.196           & 0.196           & \cb{\cg{0.194}} \\
MXX      & \cb{\cg{0.081}} & \cg{0.081}      & \cg{0.081}      & \cg{0.082}      & 0.174           & 0.174           & \cb{\cg{0.174}} & 0.174           \\
N225     & \cg{0.117}      & \cg{0.117}      & \cg{0.117}      & \cb{\cg{0.116}} & 0.186           & 0.187           & 0.186           & \cb{\cg{0.183}} \\
NSEI     & \cg{0.127}      & \cg{0.127}      & \cg{0.127}      & \cb{\cg{0.126}} & 0.166           & 0.166           & 0.166           & \cb{\cg{0.165}} \\
OSEAX    & \cg{0.337}      & \cg{0.337}      & \cg{0.337}      & \cb{\cg{0.336}} & \cg{0.264}      & 0.265           & \cg{0.264}      & \cb{\cg{0.264}} \\
RUT      & \cg{0.102}      & \cg{0.102}      & \cg{0.102}      & \cb{\cg{0.099}} & \cg{0.204}      & \cg{0.204}      & \cg{0.204}      & \cb{\cg{0.204}} \\
SPX      & 0.113           & 0.112           & 0.113           & \cb{\cg{0.110}} & 0.197           & 0.197           & 0.198           & \cb{\cg{0.195}} \\
SSEC     & 0.063           & \cb{\cg{0.063}} & 0.063           & 0.064           & 0.170           & \cb{\cg{0.170}} & \cg{0.170}      & 0.171           \\
SSMI     & \cg{0.085}      & \cg{0.085}      & \cg{0.085}      & \cb{\cg{0.084}} & \cg{0.134}      & 0.134           & \cg{0.134}      & \cb{\cg{0.133}} \\
STOXX50E & \cg{0.164}      & \cb{\cg{0.164}} & \cg{0.164}      & \cg{0.164}      & \cb{\cg{0.225}} & \cg{0.225}      & \cg{0.225}      & \cg{0.225}      \\
\bottomrule
\end{tabular}
\caption{MSE and MAE results comparisons across different GSP-HAR variants at the short-term forecasting horizon ($h=1$). Cells highlighted in gray are models selected by the MCS test at the $25\%$ significance level; numbers in blue indicate the best-performing model per index.}
\label{tab:mse_mae_laplacian_h1}
\end{table}

\begin{table}[H]
\centering
\footnotesize
\setlength{\tabcolsep}{3pt}
\renewcommand{\arraystretch}{0.95}
\begin{tabular}{l|cccc|cccc}
\toprule
         & \multicolumn{4}{c|}{MSE ($h=5$)}                     & \multicolumn{4}{c}{MAE ($h=5$)} \\
\cmidrule(lr){2-5}\cmidrule(lr){6-9}
GSP-HAR   & P               & GL              & DY-sym          & DY-asym         & P               & GL              & DY-sym          & DY-asym         \\
\midrule
AEX      & \cg{0.158}      & \cg{0.156}      & \cg{0.156}      & \cb{\cg{0.156}} & \cg{0.236}      & 0.236           & \cg{0.236}      & \cb{\cg{0.233}} \\
AORD     & \cg{0.157}      & \cg{0.156}      & \cb{\cg{0.156}} & \cg{0.157}      & \cb{\cg{0.218}} & 0.222           & \cg{0.221}      & 0.225           \\
BFX      & \cb{\cg{0.139}} & \cg{0.140}      & \cg{0.140}      & \cg{0.139}      & \cg{0.224}      & 0.225           & \cg{0.224}      & \cb{\cg{0.221}} \\
BSESN    & \cg{0.190}      & \cg{0.187}      & \cg{0.187}      & \cb{\cg{0.185}} & \cg{0.210}      & 0.211           & 0.211           & \cb{\cg{0.208}} \\
BVSP     & \cg{0.203}      & \cg{0.204}      & \cg{0.204}      & \cb{\cg{0.203}} & 0.253           & 0.250           & \cg{0.250}      & \cb{\cg{0.248}} \\
DJI      & \cg{0.207}      & \cg{0.207}      & \cg{0.207}      & \cb{\cg{0.205}} & \cb{\cg{0.262}} & \cg{0.263}      & \cg{0.264}      & \cg{0.266}      \\
FCHI     & \cg{0.160}      & \cg{0.157}      & \cg{0.157}      & \cb{\cg{0.156}} & \cg{0.247}      & 0.249           & 0.249           & \cb{\cg{0.243}} \\
FTSE     & \cg{0.265}      & \cg{0.265}      & \cg{0.265}      & \cb{\cg{0.264}} & \cg{0.273}      & 0.273           & 0.273           & \cb{\cg{0.270}} \\
GDAXI    & \cg{0.131}      & \cg{0.131}      & \cg{0.130}      & \cb{\cg{0.128}} & \cg{0.229}      & 0.230           & \cg{0.228}      & \cb{\cg{0.227}} \\
GSPTSE   & \cg{0.106}      & \cg{0.109}      & \cb{\cg{0.104}} & \cg{0.106}      & \cb{\cg{0.181}} & 0.199           & \cg{0.183}      & 0.195           \\
HSI      & \cg{0.103}      & \cb{\cg{0.102}} & \cg{0.102}      & \cg{0.104}      & \cb{\cg{0.189}} & \cb{\cg{0.189}} & \cb{\cg{0.189}} & \cb{\cg{0.189}} \\
IBEX     & \cg{0.215}      & \cb{\cg{0.212}} & \cg{0.212}      & \cg{0.213}      & 0.251           & \cg{0.247}      & 0.248           & \cb{\cg{0.246}} \\
IXIC     & \cg{0.217}      & \cb{\cg{0.217}} & \cg{0.217}      & \cg{0.217}      & \cb{\cg{0.294}} & \cb{\cg{0.294}} & \cb{\cg{0.294}} & \cb{\cg{0.294}} \\
KS11     & \cg{0.098}      & \cg{0.097}      & \cg{0.097}      & \cb{\cg{0.097}} & \cb{\cg{0.178}} & \cg{0.181}      & \cg{0.180}      & \cg{0.179}      \\
KSE      & 0.137           & 0.135           & 0.136           & \cb{\cg{0.134}} & \cg{0.230}      & \cg{0.231}      & \cg{0.230}      & \cb{\cg{0.227}} \\
MXX      & \cg{0.099}      & \cb{\cg{0.099}} & \cg{0.099}      & \cg{0.099}      & \cg{0.199}      & 0.200           & 0.200           & \cb{\cg{0.197}} \\
N225     & \cg{0.166}      & \cg{0.164}      & \cg{0.164}      & \cb{\cg{0.163}} & \cg{0.239}      & 0.240           & \cg{0.239}      & \cb{\cg{0.237}} \\
NSEI     & \cg{0.194}      & \cg{0.191}      & \cg{0.191}      & \cb{\cg{0.188}} & \cg{0.210}      & 0.211           & 0.211           & \cb{\cg{0.207}} \\
OSEAX    & \cg{0.400}      & \cg{0.400}      & \cg{0.400}      & \cb{\cg{0.399}} & 0.303           & 0.303           & 0.302           & \cb{\cg{0.299}} \\
RUT      & \cg{0.184}      & \cb{\cg{0.184}} & \cg{0.184}      & \cg{0.185}      & \cg{0.267}      & \cg{0.267}      & \cg{0.267}      & \cb{\cg{0.264}} \\
SPX      & \cg{0.221}      & \cg{0.219}      & \cg{0.219}      & \cb{\cg{0.217}} & \cg{0.284}      & \cb{\cg{0.283}} & \cb{\cg{0.283}} & \cg{0.285}      \\
SSEC     & \cg{0.095}      & \cg{0.095}      & \cb{\cg{0.095}} & \cg{0.096}      & \cg{0.212}      & 0.213           & 0.213           & \cb{\cg{0.210}} \\
SSMI     & \cg{0.145}      & 0.145           & \cg{0.145}      & \cb{\cg{0.141}} & \cb{\cg{0.177}} & \cg{0.180}      & \cg{0.180}      & \cb{\cg{0.177}} \\
STOXX50E & \cg{0.253}      & \cg{0.252}      & \cg{0.252}      & \cb{\cg{0.251}} & \cg{0.281}      & \cg{0.280}      & \cg{0.280}      & \cb{\cg{0.277}} \\
\bottomrule
\end{tabular}
\caption{MSE and MAE results comparisons across different GSP-HAR variants at the mid-term forecasting horizon ($h=5$). Cells highlighted in gray are models selected by the MCS test at the $25\%$ significance level; numbers in blue indicate the best-performing model per index.}
\label{tab:mse_mae_laplacian_h5}
\end{table}

\begin{table}[H]
\centering
\footnotesize
\setlength{\tabcolsep}{3pt}
\renewcommand{\arraystretch}{0.95}
\begin{tabular}{l|cccc|cccc}
\toprule
         & \multicolumn{4}{c|}{MSE ($h=22$)}                     & \multicolumn{4}{c}{MAE ($h=22$)} \\
\cmidrule(lr){2-5}\cmidrule(lr){6-9}
GSP-HAR   & P               & GL              & DY-sym          & DY-asym         & P               & GL              & DY-sym          & DY-asym         \\
\midrule
AEX      & \cg{0.233}      & \cg{0.234}      & \cg{0.234} & \cb{\cg{0.230}} & 0.279      & 0.278           & 0.279           & \cb{\cg{0.266}} \\
AORD     & \cb{\cg{0.237}} & \cg{0.237}      & 0.237      & 0.252           & 0.273      & \cb{\cg{0.272}} & 0.275           & 0.310           \\
BFX      & \cb{\cg{0.201}} & \cg{0.201}      & \cg{0.201} & \cg{0.202}      & \cg{0.260} & \cg{0.260}      & \cg{0.260}      & \cb{\cg{0.260}} \\
BSESN    & \cg{0.265}      & \cg{0.265}      & \cg{0.265} & \cb{\cg{0.264}} & 0.255      & \cb{\cg{0.254}} & 0.256           & 0.268           \\
BVSP     & 0.313           & 0.313           & 0.313      & \cb{\cg{0.304}} & 0.304      & 0.304           & 0.305           & \cb{\cg{0.291}} \\
DJI      & \cg{0.326}      & \cg{0.326}      & \cg{0.326} & \cb{\cg{0.321}} & \cg{0.326} & \cb{\cg{0.325}} & 0.327           & 0.334           \\
FCHI     & \cg{0.235}      & \cg{0.236}      & \cg{0.234} & \cb{\cg{0.233}} & 0.291      & 0.290           & 0.289           & \cb{\cg{0.273}} \\
FTSE     & \cg{0.358}      & 0.358           & \cg{0.358} & \cb{\cg{0.355}} & 0.313      & 0.313           & 0.314           & \cb{\cg{0.303}} \\
GDAXI    & \cg{0.198}      & \cg{0.198}      & \cg{0.198} & \cb{\cg{0.196}} & 0.261      & 0.262           & \cb{\cg{0.260}} & 0.270           \\
GSPTSE   & \cg{0.181}      & \cb{\cg{0.169}} & \cg{0.187} & \cg{0.185}      & \cg{0.249} & \cg{0.253} & \cb{\cg{0.249}}      & 0.278           \\
HSI      & \cg{0.117}      & \cg{0.117}      & \cg{0.117} & \cb{\cg{0.116}} & 0.214      & 0.214           & 0.214           & \cb{\cg{0.211}} \\
IBEX     & \cg{0.280}      & \cg{0.280}      & \cg{0.280} & \cb{\cg{0.278}} & 0.279      & \cb{\cg{0.279}} & 0.280           & 0.288           \\
IXIC     & \cb{\cg{0.314}} & \cg{0.314}      & \cg{0.314} & \cg{0.319}      & \cg{0.348} & \cg{0.348}      & \cg{0.349}      & \cb{\cg{0.339}} \\
KS11     & \cg{0.130}      & \cg{0.130}      & \cg{0.130} & \cb{\cg{0.130}} & \cg{0.221} & \cg{0.220}      & 0.221           & \cb{\cg{0.219}} \\
KSE      & \cg{0.180}      & \cg{0.180}      & \cg{0.180} & \cb{\cg{0.179}} & 0.271      & \cb{\cg{0.270}} & 0.271           & 0.274           \\
MXX      & 0.116           & 0.116           & 0.116      & \cb{\cg{0.112}} & 0.223      & 0.223           & 0.224           & \cb{\cg{0.213}} \\
N225     & \cg{0.205}      & \cg{0.205}      & \cg{0.205} & \cb{\cg{0.203}} & \cg{0.272} & \cb{\cg{0.272}} & 0.273           & 0.282           \\
NSEI     & \cg{0.271}      & \cg{0.271}      & \cg{0.271} & \cb{\cg{0.265}} & 0.258      & \cb{\cg{0.257}} & 0.258           & 0.268           \\
OSEAX    & 0.515           & 0.516           & 0.516      & \cb{\cg{0.506}} & 0.345      & 0.345           & 0.345           & \cb{\cg{0.333}} \\
RUT      & \cg{0.267}      & \cg{0.267}      & \cg{0.267} & \cb{\cg{0.266}} & 0.313      & 0.312           & 0.313           & \cb{\cg{0.305}} \\
SPX      & \cg{0.339}      & \cg{0.337}      & \cg{0.337} & \cb{\cg{0.330}} & \cg{0.347} & \cb{\cg{0.341}} & \cg{0.341}      & 0.369           \\
SSEC     & 0.118           & 0.118           & 0.118      & \cb{\cg{0.111}} & 0.250      & 0.249           & 0.250           & \cb{\cg{0.240}} \\
SSMI     & \cg{0.240}      & \cg{0.240}      & \cg{0.240} & \cb{\cg{0.235}} & 0.235      & 0.235           & \cb{\cg{0.234}} & 0.237           \\
STOXX50E & \cg{0.350}      & \cg{0.350}      & \cg{0.350} & \cb{\cg{0.346}} & \cg{0.316} & \cg{0.316}      & 0.317           & \cb{\cg{0.315}} \\
\bottomrule
\end{tabular}
\caption{MSE and MAE results comparisons across different GSP-HAR variants at the long-term forecasting horizon ($h=22$). Cells highlighted in gray are models selected by the MCS test at the $25\%$ significance level; numbers in blue indicate the best-performing model per index.}
\label{tab:mse_mae_laplacian_h22}
\end{table}

According to the tables above, the GSP-HAR model delivers consistent RV forecasting performance across different Laplacian constructions of the volatility spillover network, which underscores the robustness of the model design. Nevertheless, the GSP-HAR-DY-asym model, which is built on an asymmetric weight matrix derived from the DY framework and a magnetic Laplacian with a positive $q$ to capture directional spillovers, emerges as the preferred variant. It achieves the lowest forecast errors and remains in the MCS for more than half of the stock market indices across different evaluation criteria and forecasting horizons. While the MCS test often includes most GSP-HAR variants at the 25\% significance level due to their similar levels of loss, the GSP-HAR-DY-asym model consistently appears most frequently, highlighting its superior performance. 

This advantage likely stems from two key factors. First, the DY-magnetic-Laplacian method produces RV GSE patterns that most accurately reflect underlying market conditions. Thus, it provides graph Fourier bases that yield highly effective spectral representations of historical RV patterns for forecasting. Second, by explicitly accounting for the directional volatility transmission, the DY-magnetic-Laplacian captures the inherent imbalance between influential and impressionable stock markets, which is crucial to accurately model volatility spillovers. Therefore, the DY-magnetic-Laplacian method can be regarded as the most informative for capturing volatility interrelationships and market conditions, offering the strongest forecasting capability within the GSP-HAR framework for accurate RV prediction.

\subsection{GSP-HAR-Forecasted RV GSE as Indicators of Market Turbulence}\label{sec:Forecast_RV_Graph_Signal_Energy_Analysis}
This section examines the practical utility of the RV GSE forecasts generated by the proposed GSP-HAR model, which is identified as the most effective RV forecasting approach. Specifically, the model considered here is based on the asymmetric DY volatility network combined with the magnetic Laplacian matrix with $q>0$. The analysis proceeds by intuitively comparing the GSP-HAR-forecasted RV GSE with the ground-truth RV GSE shown in Figure \ref{graph_energy_DY_magnet}. The aim is not to evaluate whether the forecasted GSE perfectly replicates the ground-truth series. Forecasting accuracy has already been validated in Sections \ref{sec:Main_Experiment_Results} and \ref{sec:Different_L}. Rather, the focus is on assessing whether the GSP-HAR-forecasted RV GSE can function as an early-warning indicator of global market conditions.

As discussed in Section \ref{Volatility Spillover Network and Its GSE}, when the DY-magnetic-Laplacian method is applied, the RV GSE function is highly informative about market dynamics, distinguishing both turbulent and stable periods. Extending this property to forecasts, the GSE derived from GSP-HAR-forecasted RV is expected to provide early signals of potential market stress. To this end, the analysis employs long-term forecasts ($h=22$), as longer horizons are particularly valuable: unlike short- or mid-term forecasts, they allow decision-makers greater lead time to anticipate and respond to shifts in market conditions.

To visualize both the GSP-HAR-forecasted RV GSE and the ground-truth RV GSE functions, the procedure is similar to the one described in Section \ref{Volatility Spillover Network and Its GSE}. A yearly rolling-window with center date point $t_0$ and window size $2 \tau$ can be denoted as $\{\mathbf{v}_t \in \mathbb{R}^N\}_{t = t_0 - \tau}^{t_0 + \tau}$. The normalized magnetic Laplacian matrix $\mathbf{L}_{m,t_0}^{(q)}$ used for calculating the GSE is constructed from the rolling window data $\{\mathbf{v}_t\}_{t = t_0 - \tau}^{t_0 + \tau}$ under the DY-magnetic-Laplacian method. The GSP-HAR model is trained with the forecast horizon as $22$, the Laplacian matrix $\mathbf{L}_{m,t_0}^{(q)}$ and the training data $\{\mathbf{v}_t \in \mathbb{R}^N\}_{t = t_0 - \tau}^{t_0 + \tau}$. The ground-truth RV is $\overline{\mathbf{v}}_{t_0} = (\overline{v}_{t_0,1}, \overline{v}_{t_0,2}, \dots, \overline{v}_{t_0,N})^\top$, whereas the GSP-HAR-forecasted RV $\hat{\mathbf{v}}_{t_0}$ is generated by the trained GSP-HAR model with input data $\{\mathbf{v}_t \in \mathbb{R}^N\}_{t = t_0 + \tau - 22}^{t_0 + \tau}$. Use $E(\overline{\mathbf{v}}_{t_0})$ and $E(\hat{\mathbf{v}}_{t_0})$ to denote the ground-truth RV GSE and the GSP-HAR-forecasted RV GSE, respectively. Both $E(\overline{\mathbf{v}}_{t_0})$ and $E(\hat{\mathbf{v}}_{t_0})$ are calculated according to Equation \eqref{VSP_signal_energy} with $\mathbf{L}_{t_0} = \mathbf{L}_{m,t_0}^{(q)}$, and are plotted at $t = t_0$. Since $\hat{\mathbf{v}}_{t_0}$ is the forecast RV, it is expected that the $E(\hat{\mathbf{v}}_{t_0})$ should act as a leading indicator of the $E(\overline{\mathbf{v}}_{t_0})$. For instance, when $E(\hat{\mathbf{v}}_{t_0})$ rises sharply, $E(\overline{\mathbf{v}}_{t_0})$ should follow with an increase after a certain lag. The GSE of the GSP-HAR-forecasted RV and the ground-truth RV are presented in Figure \ref{graph_energy_GSP-HAR} below. They are scaled by the same factor so that the values of both functions are bounded within the interval $[0, 1]$.
\begin{figure}[H]
    \centering
    \includegraphics[width=18cm]{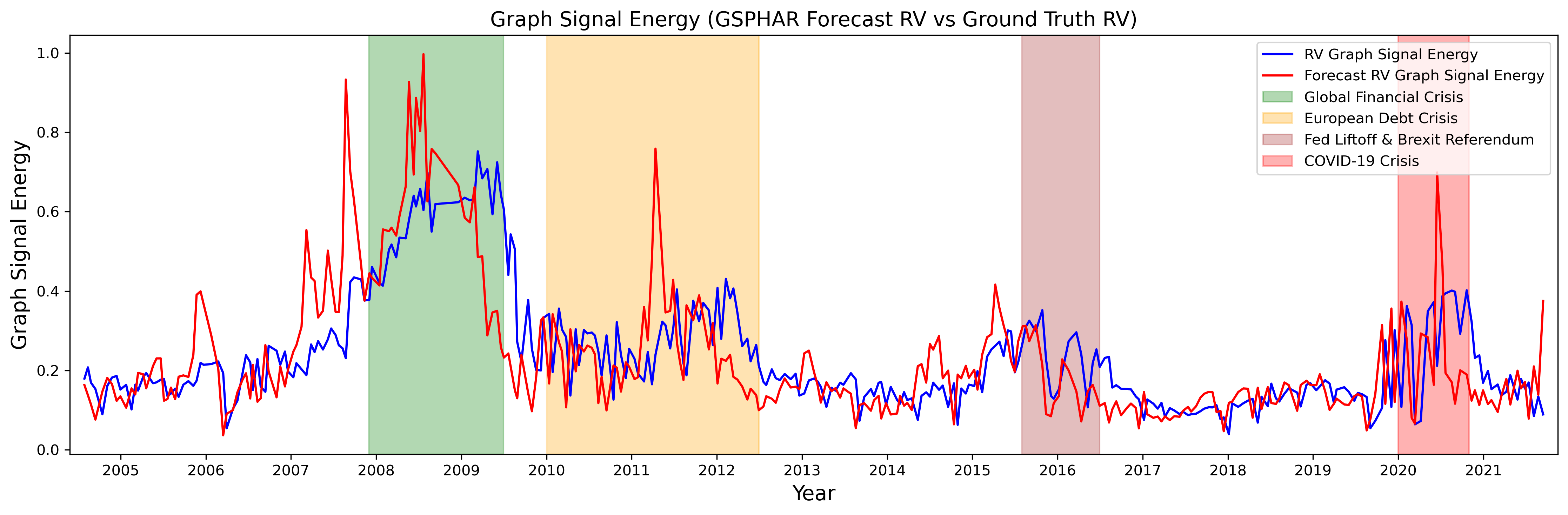}
    \caption{The GSE of the GSP-HAR-forecasted RV (red) and the ground-truth RV (blue) based on the DY-magnetic-Laplacian construction method}
    \label{graph_energy_GSP-HAR}
\end{figure}

As shown in Figure \ref{graph_energy_GSP-HAR}, the GSP-HAR-forecasted RV GSE emerges as a leading indicator of the ground-truth RV GSE, providing early-warning signals of heightened market stress. Notably, the forecasted energy function rises sharply prior to the GFC, with similar but less pronounced upward movements observed ahead of the Federal Reserve liftoff, the Brexit referendum, and the COVID-19 pandemic. Moreover, compared to the relative stability observed in tranquil periods, the forecasted energy function exhibits greater fluctuations both before and during episodes of market turbulence. These empirical patterns provide preliminary evidence that the GSP-HAR-forecasted RV GSE is both predictive and sensitive to potential financial risks. An increase in its level and volatility may signal forthcoming market instability. Thus, beyond its strength in producing accurate RV forecasts, the GSP-HAR model also demonstrates practical value as a tool for financial risk monitoring and early-warning analysis.

\section{Conclusion}
\label{sec: Conclusion}
This study examines the interconnections of global financial markets by analyzing volatility spillovers. It creatively discovers the relationship between RV GSE and the financial market conditions, which can be leveraged to measure the effectiveness of different volatility spillover network construction methods. In addition, this study proposes a novel RV forecasting framework that combines the HAR model and GSP techniques based on the magnetic Laplacian to leverage the most informative volatility spillover network for more accurate RV forecasting. Results from the empirical tests on data from $24$ global stock market indices show that, compared to baseline models, the proposed GSP-HAR model significantly improves the RV forecasting accuracy. Unlike existing RV forecast approaches, where the volatility spillover effect is regarded as additional variables representing spatial information in the HAR framework, the proposed GSP-HAR model uses the GFT technique to obtain effective RV representations in the spectral domain based on the volatility spillover network structure, and leverages the HAR model as a linear, data-driven filter to obtain the spectral volatility signal. Subsequently, the IGFT converts the filtered spectral RV signal back to the spatial domain, where a shallow neural network is applied to derive the final RV forecast in a nonlinear way. Thus, the volatility spillover effect is successfully integrated into the model's design.

In this study, the exploration of market regime identification and turbulence forecasting is limited to an intuitive, preliminary analysis based on the GSE function. While the findings suggest that the RV GSE derived from the DY-magnetic-Laplacian method can distinguish between turbulent and stable periods, and that the GSE of GSP-HAR-forecasted RV is predictive of potential market stress, these insights remain exploratory. Extending the GSP-HAR framework into more systematic applications, such as financial early-warning systems and formal market regime analysis, offers potential direction for future research.




\newpage

\newpage
\clearpage
\appendixtitleon
\appendixtitletocon
\begin{appendices}

\section{Undirected vs Directed Graphs}
\label{Undirected vs Directed Graphs}
\begin{figure}[H]
  \centering
  \begin{subfigure}[b]{0.3\textwidth}
    \centering
    \includegraphics[width=\textwidth]{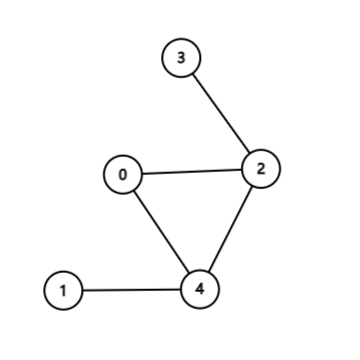}
    \caption{An example of undirected graph}
    \label{An example of undirected graph}
  \end{subfigure}
  \hspace{0.05\textwidth} 
  \begin{subfigure}[b]{0.3\textwidth}
    \centering
    \includegraphics[width=\textwidth]{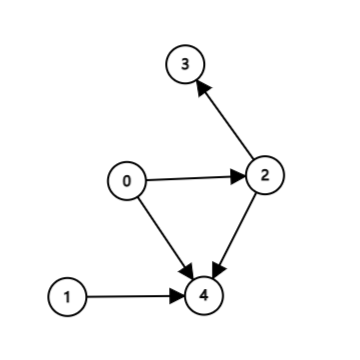}
    \caption{An example of directed graph}
    \label{An example of directed graph}
  \end{subfigure}
  \caption{Undirected and directed graphs}
  \label{Undirected and directed graphs}
\end{figure}

\section{Theorems and Corresponding Proofs}
\label{Theorems and Corresponding Proofs}
In general, suppose the graph signal $\mathbf{x} \in \mathbb{R}^N$ is defined on a $N$-node graph $\mathcal{G} = (\mathcal{V}, \mathcal{E})$, whose weight matrix and degree matrix are $\mathbf{W} \in \mathbb{R}^{N \times N}$ and $\mathbf{D} \in \mathbb{R}^{N \times N}$, respectively. The weight matrix $\mathbf{W}$ is assumed to contain non-negative values. And the degree matrix $\mathbf{D}$ is assumed to be invertible. The normalized Laplacian matrix is formulated as: $\mathbf{L} = \mathbf{I} - \mathbf{D}^{-\frac{1}{2}}\mathbf{W}\mathbf{D}^{-\frac{1}{2}}$. The normalized magnetic Laplacian $\mathbf{L}_m^{(q)}$ formulation process is described from Equation \eqref{A^s} to Equation \eqref{normalized magnetic Laplacian}.
\subsection{Non-negative GSE}
\label{Non-negative GSE}
Considering the GSE defined in Equation \eqref{graph_signal_energy}, we have the following theorem.
\begin{reptheorem}
For any given undirected and non-negative weighted graph $\mathcal{G}$ with no isolated nodes, i.e. $\mathbf{W} = \mathbf{W}^\top$, $\mathbf{W}_{ij}\geq 0$ $(i,j=1, ..., N)$, and $d_i>0$ $(i=1, 2, ..., N)$, any graph signal $\mathbf{x} \in \mathbb{R}^N$ on $\mathcal{G}$ has non-negative energy: $E(\mathbf{x}) \geq 0$.
\end{reptheorem}

\begin{proof}
According to Equation \eqref{graph_signal_energy}, $E(\mathbf{x}) = \mathbf{x}^\top \mathbf{L} \mathbf{x}$. The normalized Laplacian matrix $\mathbf{L} = \mathbf{I} - \mathbf{D}^{-\frac{1}{2}}\mathbf{W}\mathbf{D}^{-\frac{1}{2}}$. Set $\widetilde{\mathbf{W}} = \mathbf{D}^{-\frac{1}{2}}\mathbf{W}\mathbf{D}^{-\frac{1}{2}}$ and, correspondingly, $\mathbf{L} = \mathbf{I} - \widetilde{\mathbf{W}}$. According to the assumptions, $\widetilde{\mathbf{W}}$ is symmetric ($\widetilde{\mathbf{W}} = \widetilde{\mathbf{W}}^\top$) and only contains non-negative values ($\widetilde{\mathbf{W}}_{ij} \geq 0$). And it has a row sum of $1$: $\sum_{j=1}^N\widetilde{\mathbf{W}}_{ij} = 1$. Its column sum is also $1$ due to the symmetry. The GSE calculation $E(\mathbf{x}) = \mathbf{x}^\top \mathbf{L} \mathbf{x}$ can be expanded as:
\begin{equation}
    \begin{split}
        E(\mathbf{x}) =&\; \mathbf{x}^\top \mathbf{I} \mathbf{x} - \mathbf{x}^\top \widetilde{\mathbf{W}} \mathbf{x}\\
                      =&\; \sum_{i,j} \mathbf{I}_{ij}\mathbf{x}_i\mathbf{x}_j - \sum_{i,j} \widetilde{\mathbf{W}}_{ij}\mathbf{x}_i\mathbf{x}_j\\
                      =&\; \frac{1}{2}(\sum_i \mathbf{x}_i^2 + \sum_j\mathbf{x}_j^2 - 2\sum_{i,j}\widetilde{\mathbf{W}}_{ij}\mathbf{x}_i\mathbf{x}_j)\\
                      =&\; \frac{1}{2}(\sum_i \sum_j \widetilde{\mathbf{W}}_{ij} \mathbf{x}_i^2 + \sum_j \sum_i \widetilde{\mathbf{W}}_{ij} \mathbf{x}_j^2 - 2\sum_{i,j}\widetilde{\mathbf{W}}_{ij}\mathbf{x}_i\mathbf{x}_j)\\
                      =&\; \frac{1}{2}\sum_{i,j}\widetilde{\mathbf{W}}_{ij}(\mathbf{x}_i^2+\mathbf{x}_j^2-2\mathbf{x}_i\mathbf{x}_j)\\
                      =&\; \frac{1}{2}\sum_{i,j}\widetilde{\mathbf{W}}_{ij}(\mathbf{x}_i-\mathbf{x}_j)^2
    \end{split}
\end{equation}
Here, $\forall\, i,j \in \{1,\ldots,N\},\ \widetilde{\mathbf{W}}_{ij} \geq 0$, and $\forall\, \mathbf{x} \in \mathbb{R}^N,\ (\mathbf{x}_i-\mathbf{x}_j)^2 \geq 0$. Hence, $\forall\, \mathbf{x} \in \mathbb{R}^N,\ E(\mathbf{x}) = \mathbf{x}^\top \mathbf{L} \mathbf{x} \geq 0$.
\end{proof}

\textit{Remark 1:} This also indicates that, when the conditions stated at the beginning of this section are satisfied, the normalized Laplacian matrix is positive semi-definite: $\mathbf{L} \succeq 0$. Besides, when $\mathbf{W} = \mathbf{W}^\top$, both the weight matrix $\mathbf{W}$ and the degree matrix $\mathbf{D}$ are symmetric. The normalized Laplacian matrix $\mathbf{L}$ is also symmetric. In addition, the equality $E(\mathbf{x}) = 0$ is achieved when $\mathbf{x}_i = \mathbf{x}_j$, which indicates that the node feature $\mathbf{x}_i$ is constant over each graph component. A graph component is a maximal connected subgraph in which each node is reachable from others and no additional nodes are included.

\subsection{Hermitian Magnetic Laplacian}
\label{Hermitian Magnetic Laplacian}
For any directed graph $\mathcal{G}$, its weight matrix $\mathbf{W} \in \mathbb{R}^{N \times N}$ may be asymmetric: $\mathbf{W} \neq \mathbf{W}^\top$. In this case, the normalized magnetic Laplacian can be defined as $\mathbf{L}_m^{(q)} \in \mathbb{C}^{N \times N}$ by the process from Equation \eqref{A^s} to Equation \eqref{normalized magnetic Laplacian}.  

\begin{reptheorem}
For any graph $\mathcal{G}$, the normalized magnetic Laplacian $\mathbf{L}_m^{(q)} \in \mathbb{C}^{N \times N}$ is Hermitian, that is, $\mathbf{L}_m^{(q)} = (\mathbf{L}_m^{(q)})^*$.
\end{reptheorem}

\begin{proof}
To prove that $\mathbf{L}_m^{(q)} \in \mathbb{C}^{N \times N}$ is Hermitian: $\mathbf{L}_m^{(q)} = (\mathbf{L}_m^{(q)})^*$, it suffices to show the following conjugate symmetry condition: $\forall\, i,j, \,(\mathbf{L}_m^{(q)})_{ij} = \overline{(\mathbf{L}_m^{(q)})_{ji}}$. 

According to Equation \eqref{Theta}, $(\boldsymbol{\Theta}^{(q)})^\top = 2 \pi q (\mathbf{W} - \mathbf{W}^\top)^\top = 2 \pi q (\mathbf{W}^\top - \mathbf{W}) = -\boldsymbol{\Theta}^{(q)}$, which indicates $\boldsymbol{\Theta}^{(q)}_{ji} = -\boldsymbol{\Theta}^{(q)}_{ij}$. For each element in the element-wise exponential operation, $\exp(\mathrm{i} \boldsymbol{\Theta}^{(q)}_{ij}) = \cos(\boldsymbol{\Theta}^{(q)}_{ij}) + \mathrm{i} \sin(\boldsymbol{\Theta}^{(q)}_{ij})$. $\exp(\mathrm{i} \boldsymbol{\Theta}^{(q)})$ is a Hermitian matrix because:
\begin{equation}\label{Hermitian exp(iota Theta^q)}
    \begin{split}
        \exp(\mathrm{i} \boldsymbol{\Theta}^{(q)})_{ji} =&\; \exp(\mathrm{i} \boldsymbol{\Theta}^{(q)}_{ji})\\
                                                   =&\; \cos(\boldsymbol{\Theta}^{(q)}_{ji}) + \mathrm{i} \sin(\boldsymbol{\Theta}^{(q)}_{ji})\\
                                                   =&\; \cos(-\boldsymbol{\Theta}^{(q)}_{ij}) + \mathrm{i} \sin(-\boldsymbol{\Theta}^{(q)}_{ij})\\
                                                   =&\; \cos(\boldsymbol{\Theta}^{(q)}_{ij}) - \mathrm{i} \sin(\boldsymbol{\Theta}^{(q)}_{ij})\\
                                                   =&\;\overline{\exp(\mathrm{i} \boldsymbol{\Theta}^{(q)})_{ij}}.
    \end{split}
\end{equation}

Set $\widetilde{\mathbf{W}}^s = (\mathbf{D}^s)^{-\frac{1}{2}} \mathbf{W}^s (\mathbf{D}^s)^{-\frac{1}{2}}$, where $\mathbf{W}^s$ and $\mathbf{D}^s$ are calculated in Equation \eqref{A^s} and Equation \eqref{D^s}, respectively. $\widetilde{\mathbf{W}}^s$ is symmetric: $\widetilde{\mathbf{W}}^s = (\widetilde{\mathbf{W}}^s)^\top$. Set $\mathbf{M} = \exp(\mathrm{i} \boldsymbol{\Theta}^{(q)})$, where $\mathbf{M} = \text{Re}(\mathbf{M}) + \mathrm{i} \text{Im}(\mathbf{M})$, $(\text{Re}(\mathbf{M}))^\top = \text{Re}(\mathbf{M})$, and $(\text{Im}(\mathbf{M}))^\top = -\text{Im}(\mathbf{M})$ because $\exp(\mathrm{i} \boldsymbol{\Theta}^{(q)})$ is Hermitian.

According to Equation \eqref{normalized magnetic Laplacian}, each element of the normalized magnetic Laplacian is calculated as:
\begin{equation}
    \begin{split}
        (\mathbf{L}_m^{(q)})_{ij} =&\; \mathbf{I}_{ij} - (\widetilde{\mathbf{W}}^s_{ij}\odot\mathbf{M}_{ij})\\
                                              =&\; \mathbf{I}_{ij} - (\widetilde{\mathbf{W}}^s_{ij}  \text{Re}(\mathbf{M})_{ij} + \mathrm{i} \widetilde{\mathbf{W}}^s_{ij}  \text{Im}(\mathbf{M})_{ij})\\
                                              =&\; \mathbf{I}_{ij} - \widetilde{\mathbf{W}}^s_{ij}  \text{Re}(\mathbf{M})_{ij} - \mathrm{i} \widetilde{\mathbf{W}}^s_{ij}  \text{Im}(\mathbf{M})_{ij}\\
                                              =&\; \mathbf{I}_{ji} - \widetilde{\mathbf{W}}^s_{ji}  \text{Re}(\mathbf{M})_{ji} + \mathrm{i} \widetilde{\mathbf{W}}^s_{ji}  \text{Im}(\mathbf{M})_{ji}\\
                                              =&\;\overline{(\mathbf{L}_m^{(q)})_{ji}}
    \end{split}
\end{equation}

Thus, the normalized magnetic Laplacian $\mathbf{L}_m^{(q)}$ is proved to be Hermitian.
\end{proof} 

\subsection{Real, Non-negative GSE Based on Hermitian Laplacian}
\label{Real, Non-negative GSE Based on Hermitian Laplacian}
\begin{reptheorem}
The GSE based on the normalized magnetic Laplacian, calculated as $E(\mathbf{x}) = \mathbf{x}^\top \mathbf{L}_m^{(q)} \mathbf{x}$, is a real and non-negative scalar for graph signal $\mathbf{x} \in \mathbb{R}^N$.
\end{reptheorem}

\begin{proof}
As shown in Appendix \ref{Hermitian Magnetic Laplacian}, the normalized magnetic Laplacian $\mathbf{L}_m^{(q)} \in \mathbb{C}^{N \times N}$ is Hermitian. Set $\mathbf{L}_m^{(q)} = \text{Re}(\mathbf{L}_m^{(q)}) + \mathrm{i} \text{Im}(\mathbf{L}_m^{(q)})$. Correspondingly, $E(\mathbf{x}) = \mathbf{x}^\top \text{Re}(\mathbf{L}_m^{(q)}) \mathbf{x} + \mathrm{i} \mathbf{x}^\top \text{Im}(\mathbf{L}_m^{(q)}) \mathbf{x}$. Since $\mathbf{L}_m^{(q)}$ is Hermitian, $(\text{Re}(\mathbf{L}_m^{(q)}))^\top = \text{Re}(\mathbf{L}_m^{(q)})$ and $(\text{Im}(\mathbf{L}_m^{(q)}))^\top = -\text{Im}(\mathbf{L}_m^{(q)})$. Set $s = \mathbf{x}^\top \text{Im}(\mathbf{L}_m^{(q)}) \mathbf{x}$ and $s \in \mathbb{R}$.
\begin{equation}
    \begin{split}
        s =&\;s^\top \\
          =&\; (\mathbf{x}^\top \text{Im}(\mathbf{L}_m^{(q)}) \mathbf{x})^\top \\
          =&\; \mathbf{x}^\top (\text{Im}(\mathbf{L}_m^{(q)}))^\top \mathbf{x} \\
          =&\; -\mathbf{x}^\top \text{Im}(\mathbf{L}_m^{(q)}) \mathbf{x} \\
          =&\; -s \\
    \end{split}    
\end{equation}

Given $s = -s$ and $s \in \mathbb{R}$, $s = 0$. Hence, $E(\mathbf{x}) = \mathbf{x}^\top \text{Re}(\mathbf{L}_m^{(q)}) \mathbf{x} + \mathrm{i} s = \mathbf{x}^\top \text{Re}(\mathbf{L}_m^{(q)}) \mathbf{x}$, which is a real scalar. 

Similar to Appendix \ref{Hermitian Magnetic Laplacian}, set $\widetilde{\mathbf{W}}^s = (\mathbf{D}^s)^{-\frac{1}{2}} \mathbf{W}^s (\mathbf{D}^s)^{-\frac{1}{2}}$. $\widetilde{\mathbf{W}}^s$ is symmetric ($\widetilde{\mathbf{W}}^s = (\widetilde{\mathbf{W}}^s)^\top$) and only contains non-negative values ($\widetilde{\mathbf{W}}^s_{ij} \geq 0$). And it has a row sum of $1$: $\sum_{j=1}^N\widetilde{\mathbf{W}}^s_{ij} = 1$. Its column sum is also $1$ due to the symmetry.

According to Equation \eqref{normalized magnetic Laplacian}, each element of the normalized magnetic Laplacian $(\mathbf{L}_m^{(q)})_{ij} = \mathbf{I}_{ij}-\widetilde{\mathbf{W}}^s_{ij}\cos{(\boldsymbol{\Theta}_{ij})} - \mathrm{i} \widetilde{\mathbf{W}}^s_{ij} \sin{(\boldsymbol{\Theta}_{ij})}$. Hence, the element-wise representation of the real part of the normalized magnetic Laplacian is $\text{Re}(\mathbf{L}_m^{(q)})_{ij} = \mathbf{I}_{ij}-\widetilde{\mathbf{W}}^s_{ij}\cos{(\boldsymbol{\Theta}_{ij})}$. Thus, the matrix form GSE calculation, $E(\mathbf{x}) = \mathbf{x}^\top \text{Re}(\mathbf{L}_m^{(q)}) \mathbf{x}$, can be expanded as:
\begin{equation}
    \begin{split}
        E(\mathbf{x}) =&\; \sum_{i,j} \text{Re}(\mathbf{L}_m^{(q)})_{ij} \mathbf{x}_i \mathbf{x}_j \\
                      =&\; \sum_{i,j} \mathbf{I}_{ij}\mathbf{x}_i \mathbf{x}_j-\sum_{i,j}\widetilde{\mathbf{W}}^s_{ij}\mathbf{x}_i \mathbf{x}_j\cos{(\boldsymbol{\Theta}_{ij})}
    \end{split}
\end{equation}
Since $\cos{(\boldsymbol{\Theta}_{ij})} \leq 1$, the above equations can be further simplified as:
\begin{equation}
    \begin{split}
        E(\mathbf{x}) &\geq \sum_{i,j} \mathbf{I}_{ij}\mathbf{x}_i \mathbf{x}_j-\sum_{i,j}\widetilde{\mathbf{W}}^s_{ij}\mathbf{x}_i \mathbf{x}_j\\
                      =&\; \frac{1}{2}(\sum_{i}\mathbf{x}_i^2 + \sum_{j}\mathbf{x}_j^2 - 2 \sum_{i,j}\widetilde{\mathbf{W}}^s_{ij}\mathbf{x}_i \mathbf{x}_j)\\
                      =&\; \frac{1}{2}(\sum_{i}\sum_{j}\widetilde{\mathbf{W}}^s_{ij}\mathbf{x}_i^2 + \sum_{j}\sum_{i}\widetilde{\mathbf{W}}^s_{ij}\mathbf{x}_j^2 - 2 \sum_{i,j}\widetilde{\mathbf{W}}^s_{ij}\mathbf{x}_i \mathbf{x}_j) \\
                      =&\; \frac{1}{2}\sum_{i,j}\widetilde{\mathbf{W}}^s_{ij}(\mathbf{x}_i^2 + \mathbf{x}_i^2 - 2 \mathbf{x}_i \mathbf{x}_j)\\
                      =&\; \frac{1}{2}\sum_{i,j}\widetilde{\mathbf{W}}^s_{ij}(\mathbf{x}_i - \mathbf{x}_i)^2
    \end{split}
\end{equation}
Since $\forall\, i,j \in \{1,\ldots,N\},\ \widetilde{\mathbf{W}}^s_{ij} \geq 0$, and $\forall\, \mathbf{x} \in \mathbb{R}^N,\ (\mathbf{x}_i-\mathbf{x}_j)^2 \geq 0$, it follows that, $\forall\, \mathbf{x} \in \mathbb{R}^N,\ E(\mathbf{x}) = \mathbf{x}^\top \text{Re}(\mathbf{L}_m^{(q)}) \mathbf{x} \geq 0$.
\end{proof}
\textit{Remark 2:} It is important to prove that the GSE $E(\mathbf{x})$ remains a real, non-negative scalar for the graph signal $\mathbf{x} \in \mathbb{R}^N$ when the applied Laplacian matrix is Hermitian under the DY-magnetic Laplacian method. This means that the RV GSE analysis can be consistently applied to different volatility network construction methods. Thus, the resulting RV GSE functions are comparable.

In addition, it can be further proved that the normalized magnetic Laplacian matrix is positive semi-definite: $\mathbf{L}_m^{(q)} \succeq 0$. For further details, readers are referred to \citet{Zhang2021}.

\section{Volatility Spillover Network Formulation}
\label{Volatility Spillover Network Formulation}
In this section, different volatility spillover network construction methods are discussed. The dataset settings are consistent with Section \ref{Methodology and Implementation}. The given RV time series data is denoted as $\{\mathbf{v}_t \in \mathbb{R}^N \}_{t = 1}^T$, where each $\mathbf{v}_t = (v_{1,t}, v_{2,t}, \dots, v_{N,t})^\top$ represents the RV observations of $N$ assets (or markets) at time $t$. Since the volatility spillover effect refers to cross-market or cross-asset transmission, the diagonal elements of the output weight matrices of different volatility spillover construction methods are set to $0$ to exclude the own volatility component (i.e., the self-loop) from the spillover measure.

\subsection{The Pearson Correlation Matrix}
\label{The Pearson Correlation Matrix formulation}
The Pearson correlation matrix is a symmetric matrix that measures the pairwise linear dependence between variables. Each element in the Pearson correlation matrix takes a value within $[-1, 1]$. A value of $1$ indicates a perfect positive linear relationship, $-1$ indicates a perfect negative linear relationship, and $0$ indicates no linear correlation. The Pearson correlation matrix can be regarded as the weight matrix of the volatility spillover network $\mathbf{W}^{P} \in \mathbb{R}^{N \times N}$ of $N$ stock market indices and is computed as follows. 

First, for each variable $i \in \{1, \dots, N\}$, compute its sample mean and sample standard deviation in Equation \eqref{series sample mean} and Equation \eqref{series sample std}, respectively.
\begin{equation}\label{series sample mean}
    \overline{v}_i = \frac{1}{T} \sum_{t=1}^T v_{i,t}.
\end{equation}
\begin{equation}\label{series sample std}
    s_i = \sqrt{\frac{1}{T-1} \sum_{t=1}^T (v_{i,t} - \overline{v}_i)^2}.
\end{equation}
The Pearson correlation coefficient between variables $i$ and $j$ is then defined as
\begin{equation}\label{Pearson correlation coef}
    \rho_{ij} = \frac{\sum_{t=1}^T (v_{i,t} - \overline{v}_i)(v_{j,t} - \overline{v}_j)}{(T-1) \, s_i \, s_j}.
\end{equation}

Repeating this computation in Equation \eqref{Pearson correlation coef} for all $i,j \in \{1, \dots, N\}$ yields the symmetric correlation matrix $\mathbf{W}^{P} = [\rho_{ij}]$, whose diagonal entries satisfy $\mathbf{W}^P_{ii} = 0$.

\subsection{The GLASSO Precision Matrix}
\label{The GLASSO Precision Matrix formulation}
The GLASSO precision matrix is an estimate of the inverse covariance (precision) matrix that encodes the conditional dependence structure among variables, which are assumed to follow the multivariate Gaussian distribution. Here, the covariance matrix is denoted as $\boldsymbol{\Sigma}$ and the corresponding precision matrix is $\boldsymbol{\Omega}$. In the precision matrix, a zero entry $\boldsymbol{\Omega}_{ij}$ indicates conditional independence between the corresponding variables given all other variables. 

Let $\mathbf{S} \in \mathbb{R}^{N \times N}$ denote the empirical covariance matrix of the given multi-asset volatility time series $\{\mathbf{v}_t \in \mathbb{R}^N \}_{t = 1}^T$. The GLASSO estimator of the precision matrix $\widehat{\boldsymbol{\Omega}}$ solves the following optimization problem \citep{Friedman2007}:
\begin{equation}\label{GLASSO_precision_estimation}
    \widehat{\boldsymbol{\Omega}} = \arg\max_{\boldsymbol{\Omega} \succ 0} \{ \log\det(\boldsymbol{\Omega}) - \text{trace}(\mathbf{S}\boldsymbol{\Omega}) 
    - \lambda \sum_{i \neq j} |\boldsymbol{\Omega}|_{ij} \},
\end{equation}
where $\lambda > 0$ is the regularization parameter controlling the level of sparsity and $\boldsymbol{\Omega} \succ 0$ enforces positive definiteness. According to \citet{Zhang2025}, it is more beneficial to build the volatility spillover network weight matrix $\mathbf{W}^{GL} \in \mathbb{R}^{N \times N}$ based on the estimated precision matrix in the following way:
\begin{equation}\label{GLASSO_adj_mx}
   \mathbf{W}^{GL}_{ij} =
    \begin{cases}
    1, &\mbox{$\widehat{\boldsymbol{\Omega}}_{ij} \neq 0$};\\
    0, &\mbox{$\widehat{\boldsymbol{\Omega}}_{ij} = 0$ \text{or} $i=j$}.
    \end{cases}
\end{equation}

\subsection{The DY Framework}
\label{The DY Framework formulation}
The DY framework can quantify the magnitude and direction of spillovers among multiple volatility time series. It measures the spillover between time series based on the forecast error variance decomposition of a vector autoregressive (VAR) model. Suppose the given volatility time series $\{\mathbf{v}_t \in \mathbb{R}^N \}_{t = 1}^T$ is covariance stationary. A $p$-order VAR model (VAR($p$)) for these time series can be represented as:
\begin{equation}\label{VAR-p}
    \mathbf{v}_t = \sum_{i=1}^p \phi_i \mathbf{v}_{t-i} + \boldsymbol{\epsilon}_t, \;\;\boldsymbol{\epsilon}_t \sim (0, \boldsymbol{\Sigma}_{\boldsymbol{\epsilon}}),
\end{equation}
where each $\phi_i$ is a scalar coefficient. This model can be rewritten in a moving average (MA) form:
\begin{equation}\label{VAR-p-MA}
    \mathbf{v}_t = \sum_{i=1}^{\infty} \mathbf{A}_i \boldsymbol{\epsilon}_{t-i},
\end{equation}
where ${\mathbf{A}_i}$ defined recursively by $\mathbf{A}_i = \sum_{j=1}^{p} \phi_j \mathbf{A}_{i-j}$ with the assumption that $\mathbf{A}_0 = \mathbf{I}_{N \times N}$ and $\mathbf{A}_i = 0$ for $i < 0$. By performing a variance decomposition of the $h$-step-ahead forecast error, each element in the resulting matrix $\boldsymbol{\theta}^g(h)$ given by:
\begin{equation}\label{var-decomp}
    \boldsymbol{\theta}_{ij}^g(h) = \frac{\sigma_{jj}^{-1}\sum_{t=0}^{h-1}(\mathbf{e}_i^\top \mathbf{A}_h \boldsymbol{\Sigma}_{\boldsymbol{\epsilon}} \mathbf{e}_j)^2}{\sum_{t=0}^{h-1}(\mathbf{e}_i^\top \mathbf{A}_h \boldsymbol{\Sigma}_{\boldsymbol{\epsilon}} \mathbf{A}_h^\top \mathbf{e}_i)},
\end{equation}
where $h$ is the forecast horizon and $\mathbf{e}_i \in \mathbb{R}^N$ is a selection vector with $1$ in the $i$-th position and $0$ elsewhere. $\boldsymbol{\theta}_{ij}^g(h)$ is regarded as the pairwise directional volatility spillover from $j$ to $i$. To standardize matrix $\boldsymbol{\theta}^g(h)$ such that each row sums to $100$ to derive the spillover index in percentage, the elements in the normalized matrix $\widetilde{\boldsymbol{\theta}}^g(h)$ can be defined as:
\begin{equation}\label{std-var-decomp}
   \widetilde{\boldsymbol{\theta}}_{ij}^g(h) = \frac{100 \ \boldsymbol{\theta}_{ij}^g(h)}{\sum_{j=1}^N \boldsymbol{\theta}_{ij}^g(h)}.
\end{equation}

Here, in order to construct a more meaningful volatility spillover network weight matrix $\mathbf{W}^{DY}$ under the DY framework, the standardized volatility spillover index matrix $\widetilde{\boldsymbol{\theta}}^g(h)$ is transposed so that $\mathbf{W}^{DY} = \widetilde{\boldsymbol{\theta}}^g(h)^\top$ and $\mathbf{W}_{ij}^{DY}$ is the spillover index which measures the relative strength of volatility spillover from stock market index $i$ to stock market index $j$. Particularly, the diagonal elements are set to $0$: $\mathbf{W}^{DY}_{ii} = 0$. For notational simplicity, the dependence on $h$ in $\mathbf{W}^{DY}$ is omitted, although strictly it should be written as $\mathbf{W}^{DY}(h)$. In the empirical experiment section, $h$ is also the forecast horizon for forecasting RV so that the volatility spillover network construction is consistent with the RV forecast model.

Furthermore, the net pairwise volatility spillover from stock market index $i$ to stock market index $j$ is denoted as $\overline{\mathbf{W}}^{DY}_{ij}$ and can be calculated based on $\mathbf{W}^{DY}$:
\begin{equation}\label{net pairwise DY}
   \overline{\mathbf{W}}^{DY}_{ij} =
    \begin{cases}
    \mathbf{W}^{DY}_{ij} - \mathbf{W}^{DY}_{ji}, &\mbox{$\mathbf{W}^{DY}_{ij} \geq \mathbf{W}^{DY}_{ji}$};\\
    0, &\mbox{$\mathbf{W}^{DY}_{ij} < \mathbf{W}^{DY}_{ji}$}.
    \end{cases}
\end{equation}

\section{RV GSE Plots}
\label{RV GSE Plots}

\begin{figure}[H]
    \centering
    \includegraphics[width=15cm]{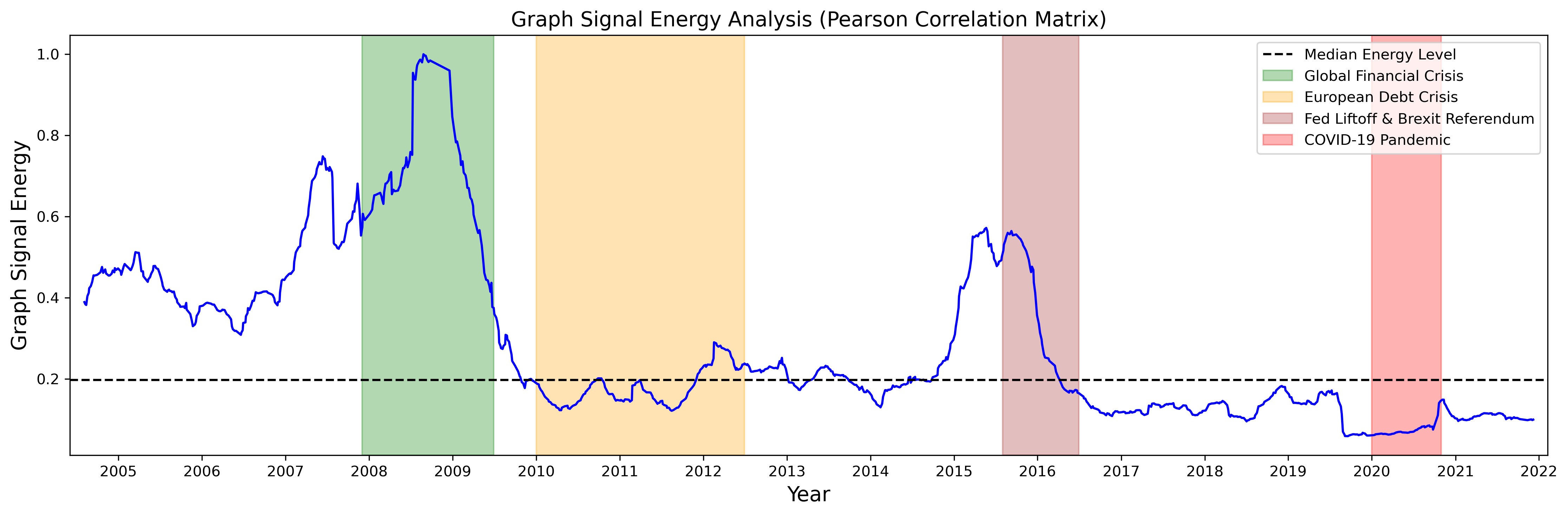}
    \caption{The RV GSE of the volatility spillover network based on the Pearson correlation matrix}
\end{figure}

\begin{figure}[H]
    \centering
    \includegraphics[width=15cm]{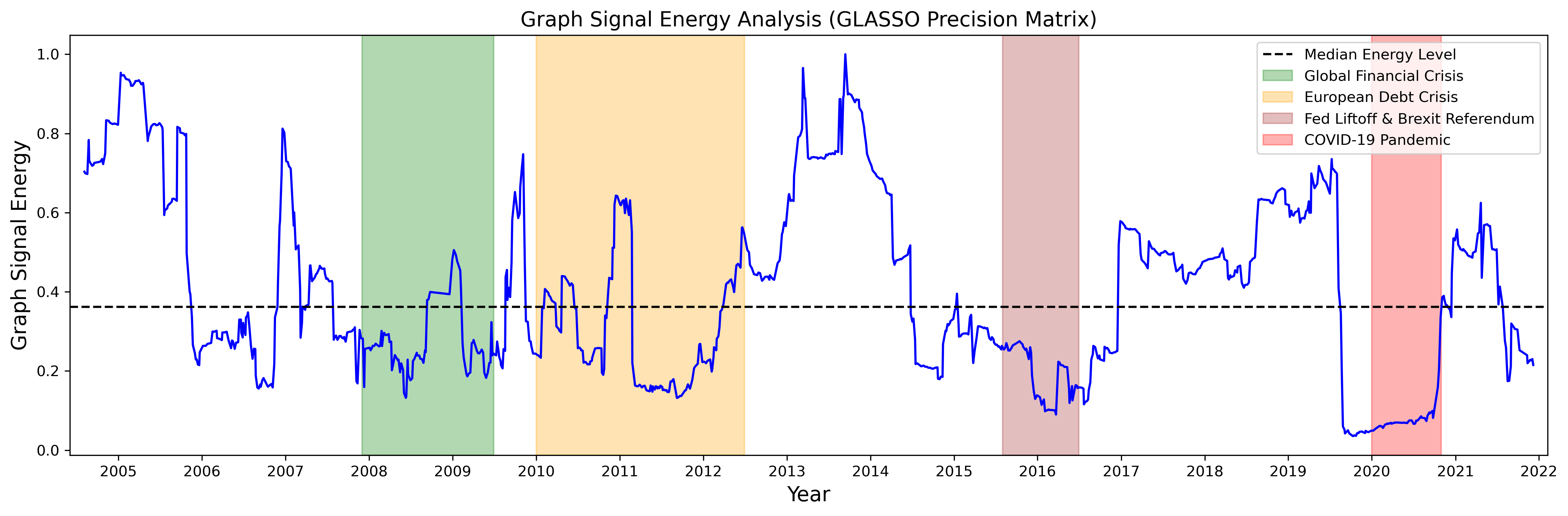}
    \caption{The RV GSE of the volatility spillover network based on the GLASSO precision matrix}
\end{figure}

\begin{figure}[H]
    \centering
    \includegraphics[width=15cm]{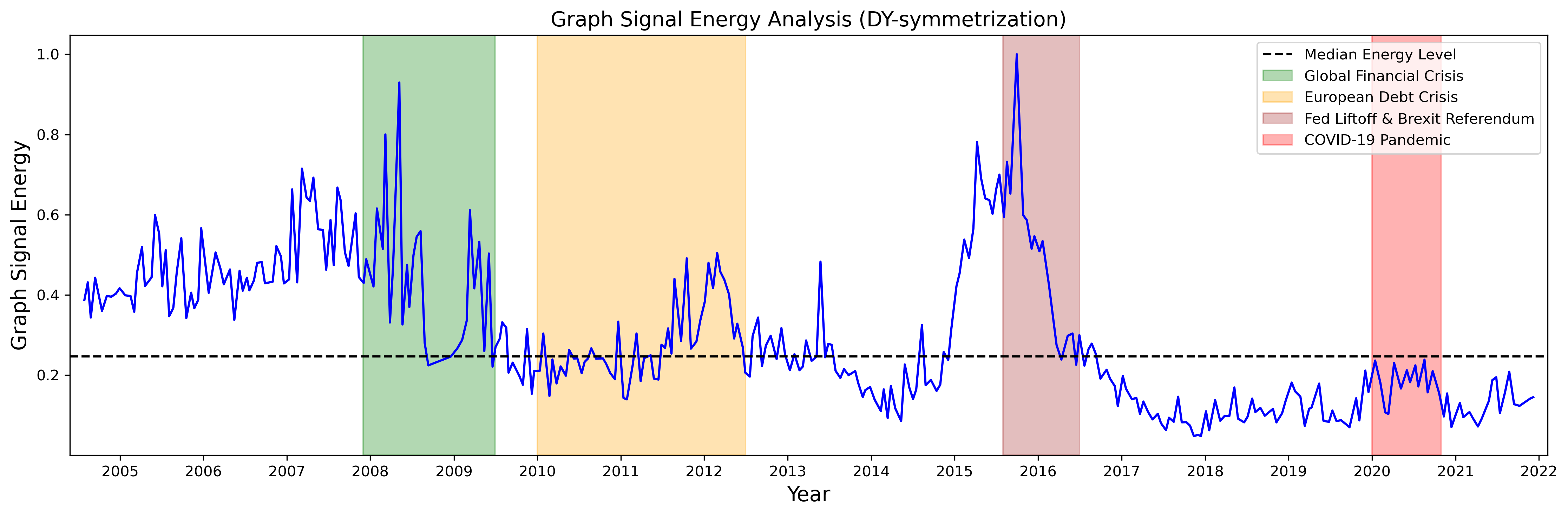}
    \caption{The RV GSE of the volatility spillover network based on the DY–symmetrization construction method}
\end{figure}

\begin{figure}[H]
    \centering
    \includegraphics[width=15cm]{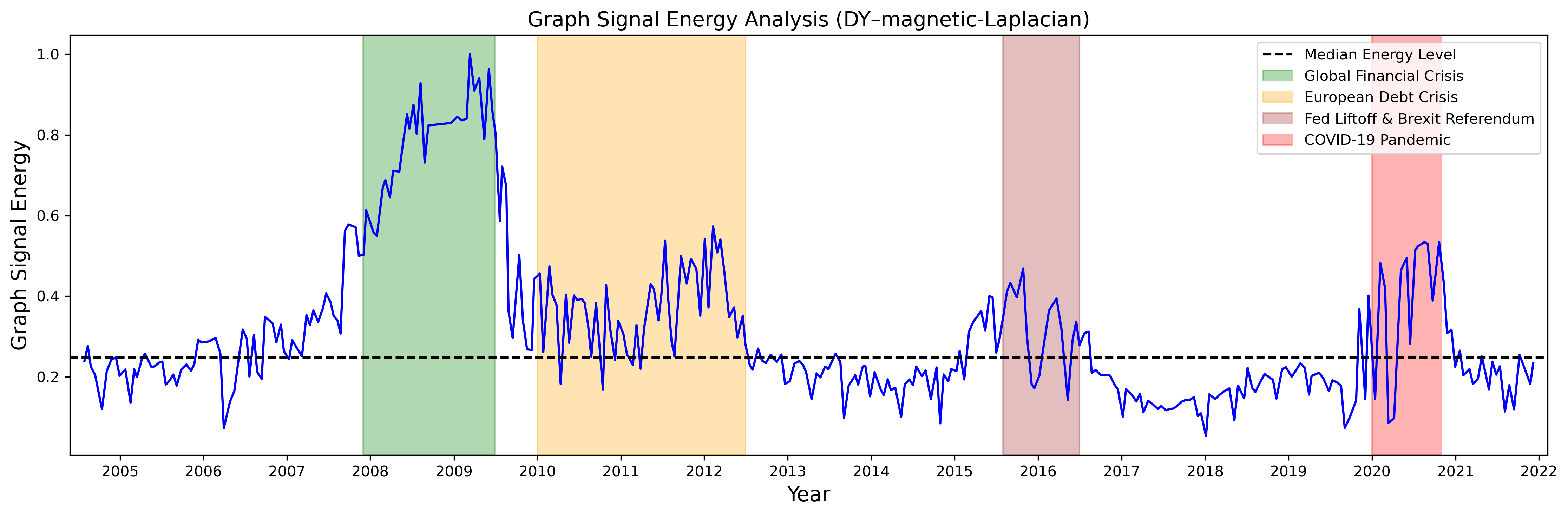}
    \caption{The RV GSE of the volatility spillover network based on the DY-magnetic-Laplacian construction method}
\end{figure}

\section{RV Time Series Data Summary Statistics}
\label{RV Time Series Data Statistics}
\begin{table}[H]
\centering
\footnotesize
\setlength{\tabcolsep}{3pt} 
\renewcommand{\arraystretch}{0.95} 
\begin{tabular}{l l c c c c}
\hline
\textbf{Index} & \textbf{Area} & \textbf{Mean} & \textbf{Std Dev} & \textbf{Skewness} & \textbf{Kurtosis} \\
\hline
AEX      & Netherlands  & 0.880 & 0.529 & 2.686 & 12.486 \\
AORD     & Australia    & 0.604 & 0.379 & 4.218 & 35.781 \\
BFX      & Belgium      & 0.816 & 0.446 & 2.604 & 12.192 \\
BSESN    & India        & 0.939 & 0.551 & 4.814 & 54.345 \\
BVSP     & Brazil       & 1.063 & 0.495 & 2.834 & 15.739 \\
DJI      & USA          & 0.808 & 0.547 & 3.201 & 19.406 \\
FCHI     & France       & 0.959 & 0.541 & 2.349 & 9.241  \\
FTSE     & UK           & 0.884 & 0.548 & 3.479 & 25.224 \\
GDAXI    & Germany      & 1.001 & 0.598 & 2.461 & 9.298  \\
GSPTSE   & Canada       & 0.669 & 0.459 & 3.518 & 21.735 \\
HSI      & China (Hong Kong) & 0.816 & 0.383 & 3.214 & 20.980 \\
IBEX     & Spain        & 1.025 & 0.557 & 2.775 & 16.836 \\
IXIC     & USA          & 0.824 & 0.477 & 2.562 & 12.042 \\
KS11     & South Korea  & 0.815 & 0.417 & 2.299 & 10.896 \\
KSE      & Pakistan     & 0.832 & 0.513 & 2.542 & 10.339 \\
MXX      & Mexico       & 0.758 & 0.409 & 3.547 & 29.078 \\
N225     & Japan        & 0.824 & 0.432 & 3.053 & 20.018 \\
NSEI     & India        & 0.946 & 0.591 & 5.272 & 70.530 \\
OSEAX    & Norway       & 0.945 & 0.588 & 5.532 & 82.454 \\
RUT      & USA          & 0.762 & 0.451 & 2.953 & 15.755 \\
SSEC     & China        & 1.085 & 0.682 & 2.445 & 9.384  \\
SSMI     & Switzerland  & 0.766 & 0.471 & 4.253 & 33.192 \\
SPX      & USA          & 0.799 & 0.546 & 2.858 & 14.293 \\
STOXX50E & Eurozone     & 1.033 & 0.611 & 2.646 & 12.626 \\

\hline
\end{tabular}
\caption{Descriptive Statistics for the RV Dataset of Various Stock Market Indices}
\label{rv_stats}
\end{table}

\end{appendices}

\newpage
\bibliographystyle{chicago}

\begin{thebibliography}{}

\bibitem[\protect\citeauthoryear{Andersen and Bollerslev}{Andersen and Bollerslev}{1998}]{Andersen1998}
Andersen, T.~G. and T.~Bollerslev (1998).
\newblock Answering the skeptics: Yes, standard volatility models do provide accurate forecasts.
\newblock {\em International Economic Review\/}~{\em 39\/}(4), 885--905.

\bibitem[\protect\citeauthoryear{BenSaïda, Litimi, and Abdallah}{BenSaïda et~al.}{2018}]{Bensaida2018}
BenSaïda, A., H.~Litimi, and O.~Abdallah (2018).
\newblock Volatility spillover shifts in global financial markets.
\newblock {\em Economic Modelling\/}~{\em 73}, 343--353.

\bibitem[\protect\citeauthoryear{Bollerslev, Hood, Huss, and Pedersen}{Bollerslev et~al.}{2018}]{Bollerslev2018}
Bollerslev, T., B.~Hood, J.~Huss, and L.~H. Pedersen (2018).
\newblock {Risk Everywhere: Modeling and Managing Volatility}.
\newblock {\em The Review of Financial Studies\/}~{\em 31\/}(7), 2729--2773.

\bibitem[\protect\citeauthoryear{Bubák, Kočenda, and Žikeš}{Bubák et~al.}{2011}]{Bubak2011}
Bubák, V., E.~Kočenda, and F.~Žikeš (2011).
\newblock Volatility transmission in emerging european foreign exchange markets.
\newblock {\em Journal of Banking \& Finance\/}~{\em 35\/}(11), 2829--2841.

\bibitem[\protect\citeauthoryear{Corsi}{Corsi}{2009}]{Corsi2009}
Corsi, F. (2009).
\newblock {A Simple Approximate Long-Memory Model of Realized Volatility}.
\newblock {\em Journal of Financial Econometrics\/}~{\em 7\/}(2), 174--196.

\bibitem[\protect\citeauthoryear{Diebold and Yilmaz}{Diebold and Yilmaz}{2009}]{Diebold2009}
Diebold, F.~X. and K.~Yilmaz (2009).
\newblock {Measuring Financial Asset Return and Volatility Spillovers, with Application to Global Equity Markets}.
\newblock {\em The Economic Journal\/}~{\em 119\/}(534), 158--171.

\bibitem[\protect\citeauthoryear{Diebold and Yilmaz}{Diebold and Yilmaz}{2012}]{Diebold2012}
Diebold, F.~X. and K.~Yilmaz (2012).
\newblock Better to give than to receive: Predictive directional measurement of volatility spillovers.
\newblock {\em International Journal of Forecasting\/}~{\em 28\/}(1), 57--66.
\newblock Special Section 1: The Predictability of Financial Markets Special Section 2: Credit Risk Modelling and Forecasting.

\bibitem[\protect\citeauthoryear{Diebold and Yılmaz}{Diebold and Yılmaz}{2014}]{Diebold2014}
Diebold, F.~X. and K.~Yılmaz (2014).
\newblock On the network topology of variance decompositions: Measuring the connectedness of financial firms.
\newblock {\em Journal of Econometrics\/}~{\em 182\/}(1), 119--134.
\newblock Causality, Prediction, and Specification Analysis: Recent Advances and Future Directions.

\bibitem[\protect\citeauthoryear{Fanuel, Ala\'{\i}z, Fern\'andez, and Suykens}{Fanuel et~al.}{2018}]{Fanuel2018}
Fanuel, M., C.~M. Ala\'{\i}z, A.~Fern\'andez, and J.~A. Suykens (2018, January).
\newblock Magnetic eigenmaps for the visualization of directed networks.
\newblock {\em Applied and Computational Harmonic Analysis\/}~{\em 44\/}(1), 189–199.

\bibitem[\protect\citeauthoryear{Forbes and Rigobon}{Forbes and Rigobon}{2002}]{Forbes2002}
Forbes, K.~J. and R.~Rigobon (2002).
\newblock No contagion, only interdependence: Measuring stock market comovements.
\newblock {\em The Journal of Finance\/}~{\em 57\/}(5), 2223--2261.

\bibitem[\protect\citeauthoryear{Friedman, Hastie, and Tibshirani}{Friedman et~al.}{2007}]{Friedman2007}
Friedman, J., T.~Hastie, and R.~Tibshirani (2007).
\newblock {Sparse inverse covariance estimation with the graphical lasso}.
\newblock {\em Biostatistics\/}~{\em 9\/}(3), 432--441.

\bibitem[\protect\citeauthoryear{Guo and Mohar}{Guo and Mohar}{2017}]{Guo2017}
Guo, K. and B.~Mohar (2017).
\newblock Hermitian adjacency matrix of digraphs and mixed graphs.
\newblock {\em Journal of Graph Theory\/}~{\em 85\/}(1), 217--248.

\bibitem[\protect\citeauthoryear{Hansen, Lunde, and Nason}{Hansen et~al.}{2011}]{Hansen2011}
Hansen, P.~R., A.~Lunde, and J.~M. Nason (2011).
\newblock The model confidence set.
\newblock {\em Econometrica\/}~{\em 79\/}(2), 453--497.

\bibitem[\protect\citeauthoryear{Kanas}{Kanas}{2000}]{Kanas2000}
Kanas, A. (2000).
\newblock Volatility spillovers between stock returns and exchange rate changes: International evidence.
\newblock {\em Journal of Business Finance \& Accounting\/}~{\em 27\/}(3-4), 447--467.

\bibitem[\protect\citeauthoryear{Li, Yu, Shahabi, and Liu}{Li et~al.}{2018}]{Li2018}
Li, Y., R.~Yu, C.~Shahabi, and Y.~Liu (2018).
\newblock Diffusion convolutional recurrent neural network: Data-driven traffic forecasting.
\newblock In {\em International Conference on Learning Representations (ICLR '18)}.

\bibitem[\protect\citeauthoryear{Liang, Wei, and Zhang}{Liang et~al.}{2020}]{Liang2020}
Liang, C., Y.~Wei, and Y.~Zhang (2020).
\newblock Is implied volatility more informative for forecasting realized volatility: An international perspective.
\newblock {\em Journal of Forecasting\/}~{\em 39\/}(8), 1253--1276.

\bibitem[\protect\citeauthoryear{Mohar}{Mohar}{2020}]{Mohar2020}
Mohar, B. (2020).
\newblock A new kind of hermitian matrices for digraphs.
\newblock {\em Linear Algebra and its Applications\/}~{\em 584}, 343--352.

\bibitem[\protect\citeauthoryear{Ortega, Frossard, Kova{\v{c}}evi{\'c}, Moura, and Vandergheynst}{Ortega et~al.}{2018}]{ortega2018graph}
Ortega, A., P.~Frossard, J.~Kova{\v{c}}evi{\'c}, J.~M. Moura, and P.~Vandergheynst (2018).
\newblock Graph signal processing: Overview, challenges, and applications.
\newblock {\em Proceedings of the IEEE\/}~{\em 106\/}(5), 808--828.

\bibitem[\protect\citeauthoryear{Patton and Sheppard}{Patton and Sheppard}{2015}]{patton2015good}
Patton, A.~J. and K.~Sheppard (2015).
\newblock Good volatility, bad volatility: Signed jumps and the persistence of volatility.
\newblock {\em Review of Economics and Statistics\/}~{\em 97\/}(3), 683--697.

\bibitem[\protect\citeauthoryear{Poon and Granger}{Poon and Granger}{2003}]{Poon2003}
Poon, S.-H. and C.~W. Granger (2003).
\newblock Forecasting volatility in financial markets: A review.
\newblock {\em Journal of Economic Literature\/}~{\em 41\/}(2), 478--539.

\bibitem[\protect\citeauthoryear{Yang and Zhou}{Yang and Zhou}{2017}]{Yang2017}
Yang, Z. and Y.~Zhou (2017).
\newblock Quantitative easing and volatility spillovers across countries and asset classes.
\newblock {\em Management Science\/}~{\em 63\/}(2), 333--354.

\bibitem[\protect\citeauthoryear{Zhang, Pu, Cucuringu, and Dong}{Zhang et~al.}{2025}]{Zhang2025}
Zhang, C., X.~Pu, M.~Cucuringu, and X.~Dong (2025).
\newblock Forecasting realized volatility with spillover effects: Perspectives from graph neural networks.
\newblock {\em International Journal of Forecasting\/}~{\em 41\/}(1), 377--397.

\bibitem[\protect\citeauthoryear{Zhang, He, Brugnone, Perlmutter, and Hirn}{Zhang et~al.}{2021}]{Zhang2021}
Zhang, X., Y.~He, N.~Brugnone, M.~Perlmutter, and M.~Hirn (2021).
\newblock Magnet: A neural network for directed graphs.
\newblock {\em Advances in Neural Information Processing Systems\/}~{\em 34}, 27003--27015.

\bibitem[\protect\citeauthoryear{Zhou, Li, Pan, and Wang}{Zhou et~al.}{2020}]{Zhou2020}
Zhou, J., D.~Li, R.~Pan, and H.~Wang (2020).
\newblock Network garch model.
\newblock {\em Statistica Sinica\/}~{\em 30\/}(4), 1723--1740.


\end{thebibliography}

\end{document}